\documentstyle{aipproc}

\begin{document}
\title{\Large\bf Neutrino Mixing and CP Violation in Matter}
\author{Zhi-zhong Xing}
\address{Sektion Physik, Universit$\it\ddot{a}$t M$\it\ddot{u}$nchen,
Theresienstrasse 37A, 80333 Munich, Germany}
\maketitle

\vspace{-0.5cm}
\begin{abstract}
Within the framework of three lepton families I present
a transparent analytical relationship between the neutrino mixing 
and CP-violating parameters in vacuum and thos in matter. Such 
a model- and parametrization-independent result will be 
particularly useful to recast the fundamental lepton flavor mixing 
matrix from the future long-baseline neutrino experiments.
\end{abstract}

Today strong evidence, that neutrinos are massive and lepton flavors 
are mixed, has been accumulated from a variety of neutrino
experiments. The mixing matrix of three different lepton
families may in general consist of non-removable complex 
phases, leading to CP or T violation. Leptonic CP violation
can manifest itself in neutrino oscillations. The best way
to observe CP- and T-violating effects is to carry out
the long-baseline neutrino experiments. 
In such experiments the earth-induced 
matter effects, which are likely to deform the neutrino oscillation behaviors 
in vacuum and fake the genuine CP-violating signals,
must be taken into account. To single out the ``true'' theory of lepton mass
generation and CP violation depends crucially upon how accurately
the fundamental parameters of lepton flavor mixing can be measured
and disentangled from the matter effects. It is therefore desirable 
to explore the most transparent analytical relationship between the 
genuine flavor mixing matrix and the matter-corrected one.

In this talk I present an exact and compact formula to describe the
matter effect on lepton flavor mixing and CP violation within the framework
of three lepton families \cite{Xing00}. 
The result is completely independent of the
specific models of neutrino masses and the specific parametrizations of
neutrino mixing. Therefore it will be particularly useful, in the long
run, to recast the fundamental flavor mixing matrix from the precise
measurements of neutrino oscillations 
in a variety of long-baseline neutrino experiments. 

In vacuum the $3\times 3$ lepton flavor mixing matrix $V$
links the neutrino mass eigenstates ($\nu_1, \nu_2, \nu_3$) to the neutrino 
flavor eigenstates ($\nu_e, \nu_\mu, \nu_\tau$).
If neutrinos are massive Dirac fermions, $V$ can be parametrized
in terms of three rotation angles and one CP-violating phase.
If neutrinos are Majorana fermions, however, two additional
CP-violating phases are in general needed to fully parametrize $V$.
The strength of CP violation in neutrino oscillations, no
matter whether neutrinos are of the Dirac or Majorana type, 
depends only upon a universal parameter $\cal J$ \cite{J}:
\begin{equation}
{\rm Im} \left (V_{\alpha i}V_{\beta j} V^*_{\alpha j}V^*_{\beta i} \right )
= {\cal J} \sum_{\gamma,k} \epsilon^{~}_{\alpha \beta \gamma} 
\epsilon^{~}_{ijk} \; ,
\end{equation}
where $(\alpha, \beta, \gamma)$ and $(i, j, k)$ run over
$(e, \mu, \tau)$ and $(1, 2, 3)$, respectively. In the specific models 
of fermion mass generation $V$ can be derived from the mass matrices 
of charged leptons and neutrinos \cite{FX99}.
To test such theoretical models one has to compare their predictions
for $V$ with the experimental data of neutrino oscillations. The latter 
may in most cases be involved in the potential matter effects and must 
be carefully handled. 

In the flavor basis where the charged lepton mass matrix $M_l$ is
diagonal (i.e., $M_l = {\rm Diag}\{ m_e, m_\mu, m_\tau \}$), the
effective Hamiltonian responsible for neutrinos propagating in
matter can be written as ${\cal H}_\nu = \Phi^{\rm m}_\nu /(2E)$
with $\Phi^{\rm m}_\nu = \Phi_\nu + \Phi_A$ and
\begin{equation}
\Phi_\nu \; = \; V \left (\matrix{
m^2_1     & 0      & 0 \cr
0         & m^2_2  & 0 \cr
0         & 0      & m^2_3 \cr} \right ) V^{\dagger} \; , ~~~~~~
\Phi_A \; = \; \left ( \matrix{
A     & 0     & 0 \cr
0     & 0     & 0 \cr
0     & 0     & 0 \cr} \right ) \; .
\end{equation}
Here $m_i$ (for $i=1, 2, 3$) denote neutrino masses,
$A = 2\sqrt{2} G_{\rm F} N_e E$ describes the charged-current 
contribution to the $\nu_e e^-$ forward scattering,
$N_e$ is the background density of electrons, and $E$ stands for
the neutrino beam energy. The neutral-current contributions, which 
are universal for $\nu_e$, $\nu_\mu$ and $\nu_\tau$ neutrinos,
lead only to an overall unobservable phase and have been neglected.
One can diagonalize $\Phi^{\rm m}_\nu$ with a unitary transformation:
${V^{\rm m}}^\dagger \Phi^{\rm m}_\nu V^{\rm m} =
{\rm Diag} \{ \lambda_1, \lambda_2, \lambda_3 \}$, where
$\lambda_i$ denote the effective mass-squared
eigenvalues of three neutrinos in matter. Explicitly we have
\begin{eqnarray}
\lambda_1 & = & m^2_1 + \frac{1}{3} x - \frac{1}{3} \sqrt{x^2 - 3y}
\left [z + \sqrt{3 \left (1-z^2 \right )} \right ] , 
\nonumber \\ 
\lambda_2 & = & m^2_1 + \frac{1}{3} x - \frac{1}{3} \sqrt{x^2 -3y}
\left [z - \sqrt{3 \left (1-z^2 \right )} \right ] , 
\nonumber \\ 
\lambda_3 & = & m^2_1 + \frac{1}{3} x + \frac{2}{3} z \sqrt{x^2 - 3y} \; ,
\end{eqnarray}
where $x$, $y$ and $z$ are given by \cite{Barger} 
\begin{eqnarray}
x & = & \Delta m^2_{21} + \Delta m^2_{31} + A \; , 
\nonumber \\
y & = & \Delta m^2_{21} \Delta m^2_{31} + A \left [ 
\Delta m^2_{21} \left ( 1 - |V_{e2}|^2 \right ) 
+ \Delta m^2_{31} \left ( 1 - |V_{e3}|^2 \right ) \right ] , 
\\
z & = & \cos \left [ \frac{1}{3} \arccos \frac{2x^3 -9xy + 27
A \Delta m^2_{21} \Delta m^2_{31} |V_{e1}|^2}
{2 \left (x^2 - 3y \right )^{3/2}} \right ] \nonumber 
\end{eqnarray}
with $\Delta m^2_{21} \equiv m^2_2 - m^2_1$ and
$\Delta m^2_{31} \equiv m^2_3 - m^2_1$. 
Note that the unitary matrix $V^{\rm m}$ is just the lepton
flavor mixing matrix in matter.
After a lengthy calculation we arrive at the elements of 
$V^{\rm m}$ as \cite{Xing00}
\begin{equation}
V^{\rm m}_{\alpha i} \; = \; \frac{N_i}{D_i} V_{\alpha i} 
+ \frac{A}{D_i} V_{e i} \left [ \left (\lambda_i - m^2_j \right )
V^*_{e k} V_{\alpha k} 
+ \left (\lambda_i - m^2_k \right ) V^*_{e j} V_{\alpha j} \right ] ,
\end{equation}
where $\alpha$ runs over $(e, \mu, \tau)$ and $(i, j, k)$ over $(1, 2, 3)$
with $i \neq j \neq k$, and
\begin{eqnarray}
N_i & = & \left (\lambda_i - m^2_j \right ) \left (\lambda_i - m^2_k \right )
- A \left [\left (\lambda_i - m^2_j \right ) |V_{e k}|^2  
+ \left (\lambda_i - m^2_k \right ) |V_{e j}|^2 \right ] ,
\nonumber \\
D^2_i & = & N^2_i + A^2 |V_{e i}|^2 \left [ 
\left (\lambda_i - m^2_j \right )^2 |V_{e k}|^2  
+ \left (\lambda_i - m^2_k \right )^2 |V_{e j}|^2 \right ] . 
\end{eqnarray}
Obviously $A=0$ leads to 
$V^{\rm m}_{\alpha i} = V_{\alpha i}$. This exact and compact formula shows 
clearly how the flavor mixing matrix in vacuum is corrected by the matter 
effects. Instructive analytical approximations can be made for Eq. (5),
once the hierarchy of neutrino masses is experimentally known or
theoretically assumed \cite{FX96}.

If leptonic CP were an exact symmetry in vacuum, the determinant of
the commutator $[\Phi_\nu, M^2_l]$ would vanish. As both $M^2_l$
and $\Phi_A$ are real diagonal matrices in the chosen flavor basis,
we find that $[\Phi^{\rm m}_\nu, M^2_l] = [\Phi_\nu, M^2_l]$
holds. Then one can derive the relationship between the CP-violating
parameter ${\cal J}$ and its counterpart in matter ${\cal J}_{\rm m}$
from the equality ${\rm Det} [\Phi^{\rm m}_\nu, M^2_l] =
{\rm Det} [\Phi_\nu, M^2_l]$. Following the calculations in
Ref. \cite{J}, we arrive at
\begin{equation}
{\cal J}_{\rm m}
\left (\lambda_1 - \lambda_2 \right ) \left (\lambda_2 - \lambda_3 \right )
\left ( \lambda_3 - \lambda_1 \right ) 
\; = \; {\cal J} \Delta m^2_{21} ~\Delta m^2_{31} ~\Delta m^2_{32} \;\; .
\end{equation}
Such an elegant relation has already been observed in Ref. \cite{Harrison}.
It indicates that the matter contamination to CP- and T-violating
observables, which must be dependent upon ${\cal J}_{\rm m}$, 
is in general unavoidable. Note that both ${\cal J}_{\rm m}$ and
$\lambda_i$ are complicated functions of the matter parameter $A$.

The results obtained above are valid for neutrinos interacting with matter.
As for antineutrinos, the matter effects arise from the charged-current 
contribution to the ``$\bar{\nu}_e e^+$ forward scattering''. 
The corresponding formulas for antineutrino mixing can straightforwardly be 
obtained from Eqs. (5) and (7) through the replacements 
$V \Longrightarrow V^*$ and $A \Longrightarrow -A$. 

The matter-corrected flavor mixing and CP-violating parameters can
be determined from a variety of long-baseline neutrino experiments. 
We calculate the conversion probabilities of 
$\nu_\alpha$ (or $\bar{\nu}_\alpha$) 
to $\nu^{~}_\beta$ (or $\bar{\nu}^{~}_\beta$) neutrinos in 
matter and obtain
\begin{eqnarray}
P_{\rm m}(\nu_\alpha \rightarrow \nu^{~}_\beta) & = &
-4 \sum_{i<j} [ {\rm Re} ( V^{\rm m}_{\alpha i} V^{\rm m}_{\beta j} 
V^{\rm m *}_{\alpha j} V^{\rm m *}_{\beta i} ) 
~ \sin^2 \Delta_{ij} ] 
+ 8 {\cal J}_{\rm m} \prod_{i<j} \sin \Delta_{ij} \; ,
\nonumber \\
P_{\rm m}(\bar{\nu}_\alpha \rightarrow \bar{\nu}^{~}_\beta) & = &
-4 \sum_{i<j} 
[ {\rm Re} ( \tilde{V}^{\rm m}_{\alpha i} \tilde{V}^{\rm m}_{\beta j} 
\tilde{V}^{\rm m *}_{\alpha j} \tilde{V}^{\rm m *}_{\beta i} ) 
~ \sin^2 \tilde{\Delta}_{ij} ] 
- 8 \tilde{\cal J}_{\rm m} \prod_{i<j} \sin 
\tilde{\Delta}_{ij} \; ,
\end{eqnarray}
where $(\alpha, \beta)$ run over $(e, \mu)$, $(\mu, \tau)$ or $(\tau, e)$;
$\tilde{V}_{\alpha i}(A) \equiv V_{\alpha i}(-A)$,
$\tilde{\Delta}_{ij}(A) \equiv \Delta_{ij}(-A)$, and $\tilde{\cal J}_{\rm m}(A)
\equiv {\cal J}_{\rm m}(-A)$; 
and $\Delta_{ij} \equiv 1.27 (\lambda_i - \lambda_j)L/E$ with $L$ the
distance between the production and interaction points of $\nu_\alpha$
(in unit of km) and $E$ the neutrino beam energy (in unit of GeV). 
$P_{\rm m}(\nu^{~}_\beta \rightarrow \nu_\alpha)$ and
$P_{\rm m}(\bar{\nu}^{~}_\beta \rightarrow \bar{\nu}_\alpha)$ 
can be read off from Eq. (8) with the replacements ${\cal J}_{\rm m}
\Longrightarrow -{\cal J}_{\rm m}$ and $\tilde{\cal J}_{\rm m}
\Longrightarrow -\tilde{\cal J}_{\rm m}$, respectively. 


Let me give a numerical illustration of the matter-induced 
corrections to the lepton flavor mixing matrix in vacuum.
The elements of $V^{\rm m}$, except the unknown Majorana phases,
can be completely determined by four rephasing-invariant quantities
(e.g., four independent $|V^{\rm m}_{\alpha i}|$ or three independent 
$|V^{\rm m}_{\alpha i}|$ plus ${\cal J}_{\rm m}$). As the solar
and atmospheric neutrino oscillations in vacuum are essentially 
associated with the elements in the first row and the third column of $V$,
it is favored to choose $|V_{e1}|$, $|V_{e2}|$, 
$|V_{\mu 3}|$ and ${\cal J}$ as the four basic parameters. 
To be specific we 
take $|V_{e1}| = 0.816$, $|V_{e2}| = 0.571$, $|V_{\mu 3}| = 0.640$,
and ${\cal J} = \pm 0.020$ for neutrinos and antineutrinos.
Such a choice is consistent with the CHOOZ experiment \cite{CHOOZ}, the
large-angle MSW solution to the solar neutrino problem, 
and a nearly maximal mixing in the atmospheric neutrino oscillation \cite{SK}. 
The relevant neutrino mass-squared differences are typically taken to be
$\Delta m^2_{21} = 5 \cdot 10^{-5} ~ {\rm eV}^2$ and
$\Delta m^2_{31} = 3 \cdot 10^{-3} ~ {\rm eV}^2$.
Then we compute the ratios $|V^{\rm m}_{\alpha i}|/|V_{\alpha i}|$ and
${\cal J}_{\rm m}/{\cal J}$ as functions of the matter parameter $A$
in the range $10^{-7} ~ {\rm eV}^2 \leq A \leq 10^{-2} ~ {\rm eV}^2$. The 
numerical results are shown in Figs. 1 to 4.
\begin{figure}
\setlength{\unitlength}{0.240900pt}
\ifx\plotpoint\undefined\newsavebox{\plotpoint}\fi
\sbox{\plotpoint}{\rule[-0.200pt]{0.400pt}{0.400pt}}%
\begin{picture}(1200,900)(-300,0)
\font\gnuplot=cmr10 at 10pt
\gnuplot
\sbox{\plotpoint}{\rule[-0.200pt]{0.400pt}{0.400pt}}%
\put(161.0,123.0){\rule[-0.200pt]{4.818pt}{0.400pt}}
\put(141,123){\makebox(0,0)[r]{0}}
\put(1119.0,123.0){\rule[-0.200pt]{4.818pt}{0.400pt}}
\put(161.0,228.0){\rule[-0.200pt]{4.818pt}{0.400pt}}
\put(141,228){\makebox(0,0)[r]{.2}}
\put(1119.0,228.0){\rule[-0.200pt]{4.818pt}{0.400pt}}
\put(161.0,334.0){\rule[-0.200pt]{4.818pt}{0.400pt}}
\put(141,334){\makebox(0,0)[r]{.4}}
\put(1119.0,334.0){\rule[-0.200pt]{4.818pt}{0.400pt}}
\put(161.0,439.0){\rule[-0.200pt]{4.818pt}{0.400pt}}
\put(141,439){\makebox(0,0)[r]{.6}}
\put(1119.0,439.0){\rule[-0.200pt]{4.818pt}{0.400pt}}
\put(161.0,544.0){\rule[-0.200pt]{4.818pt}{0.400pt}}
\put(141,544){\makebox(0,0)[r]{.8}}
\put(1119.0,544.0){\rule[-0.200pt]{4.818pt}{0.400pt}}
\put(161.0,649.0){\rule[-0.200pt]{4.818pt}{0.400pt}}
\put(141,649){\makebox(0,0)[r]{1}}
\put(1119.0,649.0){\rule[-0.200pt]{4.818pt}{0.400pt}}
\put(161.0,755.0){\rule[-0.200pt]{4.818pt}{0.400pt}}
\put(141,755){\makebox(0,0)[r]{1.2}}
\put(1119.0,755.0){\rule[-0.200pt]{4.818pt}{0.400pt}}
\put(161.0,860.0){\rule[-0.200pt]{4.818pt}{0.400pt}}
\put(141,860){\makebox(0,0)[r]{1.4}}
\put(1119.0,860.0){\rule[-0.200pt]{4.818pt}{0.400pt}}
\put(161.0,123.0){\rule[-0.200pt]{0.400pt}{4.818pt}}
\put(161,82){\makebox(0,0){10$^{-7}$}}
\put(161.0,840.0){\rule[-0.200pt]{0.400pt}{4.818pt}}
\put(220.0,123.0){\rule[-0.200pt]{0.400pt}{2.409pt}}
\put(220.0,850.0){\rule[-0.200pt]{0.400pt}{2.409pt}}
\put(298.0,123.0){\rule[-0.200pt]{0.400pt}{2.409pt}}
\put(298.0,850.0){\rule[-0.200pt]{0.400pt}{2.409pt}}
\put(338.0,123.0){\rule[-0.200pt]{0.400pt}{2.409pt}}
\put(338.0,850.0){\rule[-0.200pt]{0.400pt}{2.409pt}}
\put(357.0,123.0){\rule[-0.200pt]{0.400pt}{4.818pt}}
\put(357,82){\makebox(0,0){10$^{-6}$}}
\put(357.0,840.0){\rule[-0.200pt]{0.400pt}{4.818pt}}
\put(415.0,123.0){\rule[-0.200pt]{0.400pt}{2.409pt}}
\put(415.0,850.0){\rule[-0.200pt]{0.400pt}{2.409pt}}
\put(493.0,123.0){\rule[-0.200pt]{0.400pt}{2.409pt}}
\put(493.0,850.0){\rule[-0.200pt]{0.400pt}{2.409pt}}
\put(533.0,123.0){\rule[-0.200pt]{0.400pt}{2.409pt}}
\put(533.0,850.0){\rule[-0.200pt]{0.400pt}{2.409pt}}
\put(552.0,123.0){\rule[-0.200pt]{0.400pt}{4.818pt}}
\put(552,82){\makebox(0,0){10$^{-5}$}}
\put(552.0,840.0){\rule[-0.200pt]{0.400pt}{4.818pt}}
\put(611.0,123.0){\rule[-0.200pt]{0.400pt}{2.409pt}}
\put(611.0,850.0){\rule[-0.200pt]{0.400pt}{2.409pt}}
\put(689.0,123.0){\rule[-0.200pt]{0.400pt}{2.409pt}}
\put(689.0,850.0){\rule[-0.200pt]{0.400pt}{2.409pt}}
\put(729.0,123.0){\rule[-0.200pt]{0.400pt}{2.409pt}}
\put(729.0,850.0){\rule[-0.200pt]{0.400pt}{2.409pt}}
\put(748.0,123.0){\rule[-0.200pt]{0.400pt}{4.818pt}}
\put(748,82){\makebox(0,0){10$^{-4}$}}
\put(748.0,840.0){\rule[-0.200pt]{0.400pt}{4.818pt}}
\put(807.0,123.0){\rule[-0.200pt]{0.400pt}{2.409pt}}
\put(807.0,850.0){\rule[-0.200pt]{0.400pt}{2.409pt}}
\put(885.0,123.0){\rule[-0.200pt]{0.400pt}{2.409pt}}
\put(885.0,850.0){\rule[-0.200pt]{0.400pt}{2.409pt}}
\put(924.0,123.0){\rule[-0.200pt]{0.400pt}{2.409pt}}
\put(924.0,850.0){\rule[-0.200pt]{0.400pt}{2.409pt}}
\put(943.0,123.0){\rule[-0.200pt]{0.400pt}{4.818pt}}
\put(943,82){\makebox(0,0){10$^{-3}$}}
\put(943.0,840.0){\rule[-0.200pt]{0.400pt}{4.818pt}}
\put(1002.0,123.0){\rule[-0.200pt]{0.400pt}{2.409pt}}
\put(1002.0,850.0){\rule[-0.200pt]{0.400pt}{2.409pt}}
\put(1080.0,123.0){\rule[-0.200pt]{0.400pt}{2.409pt}}
\put(1080.0,850.0){\rule[-0.200pt]{0.400pt}{2.409pt}}
\put(1120.0,123.0){\rule[-0.200pt]{0.400pt}{2.409pt}}
\put(1120.0,850.0){\rule[-0.200pt]{0.400pt}{2.409pt}}
\put(1139.0,123.0){\rule[-0.200pt]{0.400pt}{4.818pt}}
\put(1139,82){\makebox(0,0){10$^{-2}$}}
\put(1139.0,840.0){\rule[-0.200pt]{0.400pt}{4.818pt}}
\put(161.0,123.0){\rule[-0.200pt]{235.600pt}{0.400pt}}
\put(1139.0,123.0){\rule[-0.200pt]{0.400pt}{177.543pt}}
\put(161.0,860.0){\rule[-0.200pt]{235.600pt}{0.400pt}}
\put(20,491){\makebox(0,0){$\displaystyle\frac{|V^{\rm m}_{e1}|}{|V_{e1}|}$}}
\put(650,3){\makebox(0,0){$A$ (eV$^2$)}}
\put(610,490){\makebox(0,0){$\nu$}}
\put(900,720){\makebox(0,0){$\overline{\nu}$}}
\put(350,220){\makebox(0,0){$|V_{e1}| = 0.816$}}
\put(161.0,123.0){\rule[-0.200pt]{0.400pt}{177.543pt}}
\put(161,649){\usebox{\plotpoint}}
\put(244,647.67){\rule{3.854pt}{0.400pt}}
\multiput(244.00,648.17)(8.000,-1.000){2}{\rule{1.927pt}{0.400pt}}
\put(161.0,649.0){\rule[-0.200pt]{19.995pt}{0.400pt}}
\put(293,646.67){\rule{4.095pt}{0.400pt}}
\multiput(293.00,647.17)(8.500,-1.000){2}{\rule{2.048pt}{0.400pt}}
\put(260.0,648.0){\rule[-0.200pt]{7.950pt}{0.400pt}}
\put(326,645.67){\rule{4.095pt}{0.400pt}}
\multiput(326.00,646.17)(8.500,-1.000){2}{\rule{2.048pt}{0.400pt}}
\put(310.0,647.0){\rule[-0.200pt]{3.854pt}{0.400pt}}
\put(360,644.67){\rule{3.854pt}{0.400pt}}
\multiput(360.00,645.17)(8.000,-1.000){2}{\rule{1.927pt}{0.400pt}}
\put(376,643.67){\rule{4.095pt}{0.400pt}}
\multiput(376.00,644.17)(8.500,-1.000){2}{\rule{2.048pt}{0.400pt}}
\put(393,642.67){\rule{3.854pt}{0.400pt}}
\multiput(393.00,643.17)(8.000,-1.000){2}{\rule{1.927pt}{0.400pt}}
\put(409,641.17){\rule{3.500pt}{0.400pt}}
\multiput(409.00,642.17)(9.736,-2.000){2}{\rule{1.750pt}{0.400pt}}
\put(426,639.67){\rule{3.854pt}{0.400pt}}
\multiput(426.00,640.17)(8.000,-1.000){2}{\rule{1.927pt}{0.400pt}}
\multiput(442.00,638.95)(3.588,-0.447){3}{\rule{2.367pt}{0.108pt}}
\multiput(442.00,639.17)(12.088,-3.000){2}{\rule{1.183pt}{0.400pt}}
\put(459,635.17){\rule{3.300pt}{0.400pt}}
\multiput(459.00,636.17)(9.151,-2.000){2}{\rule{1.650pt}{0.400pt}}
\multiput(475.00,633.94)(2.382,-0.468){5}{\rule{1.800pt}{0.113pt}}
\multiput(475.00,634.17)(13.264,-4.000){2}{\rule{0.900pt}{0.400pt}}
\multiput(492.00,629.94)(2.236,-0.468){5}{\rule{1.700pt}{0.113pt}}
\multiput(492.00,630.17)(12.472,-4.000){2}{\rule{0.850pt}{0.400pt}}
\multiput(508.00,625.93)(1.823,-0.477){7}{\rule{1.460pt}{0.115pt}}
\multiput(508.00,626.17)(13.970,-5.000){2}{\rule{0.730pt}{0.400pt}}
\multiput(525.00,620.93)(1.395,-0.482){9}{\rule{1.167pt}{0.116pt}}
\multiput(525.00,621.17)(13.579,-6.000){2}{\rule{0.583pt}{0.400pt}}
\multiput(541.00,614.93)(0.961,-0.489){15}{\rule{0.856pt}{0.118pt}}
\multiput(541.00,615.17)(15.224,-9.000){2}{\rule{0.428pt}{0.400pt}}
\multiput(558.00,605.92)(0.860,-0.491){17}{\rule{0.780pt}{0.118pt}}
\multiput(558.00,606.17)(15.381,-10.000){2}{\rule{0.390pt}{0.400pt}}
\multiput(575.00,595.92)(0.616,-0.493){23}{\rule{0.592pt}{0.119pt}}
\multiput(575.00,596.17)(14.771,-13.000){2}{\rule{0.296pt}{0.400pt}}
\multiput(591.00,582.92)(0.529,-0.494){29}{\rule{0.525pt}{0.119pt}}
\multiput(591.00,583.17)(15.910,-16.000){2}{\rule{0.263pt}{0.400pt}}
\multiput(608.58,565.41)(0.494,-0.657){29}{\rule{0.119pt}{0.625pt}}
\multiput(607.17,566.70)(16.000,-19.703){2}{\rule{0.400pt}{0.313pt}}
\multiput(624.58,544.14)(0.495,-0.738){31}{\rule{0.119pt}{0.688pt}}
\multiput(623.17,545.57)(17.000,-23.572){2}{\rule{0.400pt}{0.344pt}}
\multiput(641.58,518.47)(0.494,-0.945){29}{\rule{0.119pt}{0.850pt}}
\multiput(640.17,520.24)(16.000,-28.236){2}{\rule{0.400pt}{0.425pt}}
\multiput(657.58,488.07)(0.495,-1.069){31}{\rule{0.119pt}{0.947pt}}
\multiput(656.17,490.03)(17.000,-34.034){2}{\rule{0.400pt}{0.474pt}}
\multiput(674.58,451.64)(0.494,-1.201){29}{\rule{0.119pt}{1.050pt}}
\multiput(673.17,453.82)(16.000,-35.821){2}{\rule{0.400pt}{0.525pt}}
\multiput(690.58,413.68)(0.495,-1.189){31}{\rule{0.119pt}{1.041pt}}
\multiput(689.17,415.84)(17.000,-37.839){2}{\rule{0.400pt}{0.521pt}}
\multiput(707.58,373.54)(0.494,-1.233){29}{\rule{0.119pt}{1.075pt}}
\multiput(706.17,375.77)(16.000,-36.769){2}{\rule{0.400pt}{0.538pt}}
\multiput(723.58,335.07)(0.495,-1.069){31}{\rule{0.119pt}{0.947pt}}
\multiput(722.17,337.03)(17.000,-34.034){2}{\rule{0.400pt}{0.474pt}}
\multiput(740.58,299.56)(0.495,-0.919){31}{\rule{0.119pt}{0.829pt}}
\multiput(739.17,301.28)(17.000,-29.279){2}{\rule{0.400pt}{0.415pt}}
\multiput(757.58,268.78)(0.494,-0.849){29}{\rule{0.119pt}{0.775pt}}
\multiput(756.17,270.39)(16.000,-25.391){2}{\rule{0.400pt}{0.388pt}}
\multiput(773.58,242.44)(0.495,-0.648){31}{\rule{0.119pt}{0.618pt}}
\multiput(772.17,243.72)(17.000,-20.718){2}{\rule{0.400pt}{0.309pt}}
\multiput(790.58,220.72)(0.494,-0.561){29}{\rule{0.119pt}{0.550pt}}
\multiput(789.17,221.86)(16.000,-16.858){2}{\rule{0.400pt}{0.275pt}}
\multiput(806.00,203.92)(0.566,-0.494){27}{\rule{0.553pt}{0.119pt}}
\multiput(806.00,204.17)(15.852,-15.000){2}{\rule{0.277pt}{0.400pt}}
\multiput(823.00,188.92)(0.616,-0.493){23}{\rule{0.592pt}{0.119pt}}
\multiput(823.00,189.17)(14.771,-13.000){2}{\rule{0.296pt}{0.400pt}}
\multiput(839.00,175.93)(0.961,-0.489){15}{\rule{0.856pt}{0.118pt}}
\multiput(839.00,176.17)(15.224,-9.000){2}{\rule{0.428pt}{0.400pt}}
\multiput(856.00,166.93)(0.902,-0.489){15}{\rule{0.811pt}{0.118pt}}
\multiput(856.00,167.17)(14.316,-9.000){2}{\rule{0.406pt}{0.400pt}}
\multiput(872.00,157.93)(1.485,-0.482){9}{\rule{1.233pt}{0.116pt}}
\multiput(872.00,158.17)(14.440,-6.000){2}{\rule{0.617pt}{0.400pt}}
\multiput(889.00,151.93)(1.395,-0.482){9}{\rule{1.167pt}{0.116pt}}
\multiput(889.00,152.17)(13.579,-6.000){2}{\rule{0.583pt}{0.400pt}}
\multiput(905.00,145.94)(2.382,-0.468){5}{\rule{1.800pt}{0.113pt}}
\multiput(905.00,146.17)(13.264,-4.000){2}{\rule{0.900pt}{0.400pt}}
\multiput(922.00,141.94)(2.382,-0.468){5}{\rule{1.800pt}{0.113pt}}
\multiput(922.00,142.17)(13.264,-4.000){2}{\rule{0.900pt}{0.400pt}}
\multiput(939.00,137.95)(3.365,-0.447){3}{\rule{2.233pt}{0.108pt}}
\multiput(939.00,138.17)(11.365,-3.000){2}{\rule{1.117pt}{0.400pt}}
\put(955,134.17){\rule{3.500pt}{0.400pt}}
\multiput(955.00,135.17)(9.736,-2.000){2}{\rule{1.750pt}{0.400pt}}
\put(972,132.17){\rule{3.300pt}{0.400pt}}
\multiput(972.00,133.17)(9.151,-2.000){2}{\rule{1.650pt}{0.400pt}}
\put(988,130.17){\rule{3.500pt}{0.400pt}}
\multiput(988.00,131.17)(9.736,-2.000){2}{\rule{1.750pt}{0.400pt}}
\put(1005,128.67){\rule{3.854pt}{0.400pt}}
\multiput(1005.00,129.17)(8.000,-1.000){2}{\rule{1.927pt}{0.400pt}}
\put(1021,127.67){\rule{4.095pt}{0.400pt}}
\multiput(1021.00,128.17)(8.500,-1.000){2}{\rule{2.048pt}{0.400pt}}
\put(1038,126.67){\rule{3.854pt}{0.400pt}}
\multiput(1038.00,127.17)(8.000,-1.000){2}{\rule{1.927pt}{0.400pt}}
\put(1054,125.67){\rule{4.095pt}{0.400pt}}
\multiput(1054.00,126.17)(8.500,-1.000){2}{\rule{2.048pt}{0.400pt}}
\put(343.0,646.0){\rule[-0.200pt]{4.095pt}{0.400pt}}
\put(1087,124.67){\rule{4.095pt}{0.400pt}}
\multiput(1087.00,125.17)(8.500,-1.000){2}{\rule{2.048pt}{0.400pt}}
\put(1071.0,126.0){\rule[-0.200pt]{3.854pt}{0.400pt}}
\put(1104.0,125.0){\rule[-0.200pt]{7.950pt}{0.400pt}}
\put(161,650){\usebox{\plotpoint}}
\put(244,649.67){\rule{3.854pt}{0.400pt}}
\multiput(244.00,649.17)(8.000,1.000){2}{\rule{1.927pt}{0.400pt}}
\put(161.0,650.0){\rule[-0.200pt]{19.995pt}{0.400pt}}
\put(310,650.67){\rule{3.854pt}{0.400pt}}
\multiput(310.00,650.17)(8.000,1.000){2}{\rule{1.927pt}{0.400pt}}
\put(260.0,651.0){\rule[-0.200pt]{12.045pt}{0.400pt}}
\put(343,651.67){\rule{4.095pt}{0.400pt}}
\multiput(343.00,651.17)(8.500,1.000){2}{\rule{2.048pt}{0.400pt}}
\put(360,652.67){\rule{3.854pt}{0.400pt}}
\multiput(360.00,652.17)(8.000,1.000){2}{\rule{1.927pt}{0.400pt}}
\put(376,653.67){\rule{4.095pt}{0.400pt}}
\multiput(376.00,653.17)(8.500,1.000){2}{\rule{2.048pt}{0.400pt}}
\put(393,654.67){\rule{3.854pt}{0.400pt}}
\multiput(393.00,654.17)(8.000,1.000){2}{\rule{1.927pt}{0.400pt}}
\put(409,655.67){\rule{4.095pt}{0.400pt}}
\multiput(409.00,655.17)(8.500,1.000){2}{\rule{2.048pt}{0.400pt}}
\put(426,657.17){\rule{3.300pt}{0.400pt}}
\multiput(426.00,656.17)(9.151,2.000){2}{\rule{1.650pt}{0.400pt}}
\put(442,658.67){\rule{4.095pt}{0.400pt}}
\multiput(442.00,658.17)(8.500,1.000){2}{\rule{2.048pt}{0.400pt}}
\multiput(459.00,660.61)(3.365,0.447){3}{\rule{2.233pt}{0.108pt}}
\multiput(459.00,659.17)(11.365,3.000){2}{\rule{1.117pt}{0.400pt}}
\put(475,663.17){\rule{3.500pt}{0.400pt}}
\multiput(475.00,662.17)(9.736,2.000){2}{\rule{1.750pt}{0.400pt}}
\multiput(492.00,665.61)(3.365,0.447){3}{\rule{2.233pt}{0.108pt}}
\multiput(492.00,664.17)(11.365,3.000){2}{\rule{1.117pt}{0.400pt}}
\multiput(508.00,668.60)(2.382,0.468){5}{\rule{1.800pt}{0.113pt}}
\multiput(508.00,667.17)(13.264,4.000){2}{\rule{0.900pt}{0.400pt}}
\multiput(525.00,672.60)(2.236,0.468){5}{\rule{1.700pt}{0.113pt}}
\multiput(525.00,671.17)(12.472,4.000){2}{\rule{0.850pt}{0.400pt}}
\multiput(541.00,676.59)(1.823,0.477){7}{\rule{1.460pt}{0.115pt}}
\multiput(541.00,675.17)(13.970,5.000){2}{\rule{0.730pt}{0.400pt}}
\multiput(558.00,681.59)(1.485,0.482){9}{\rule{1.233pt}{0.116pt}}
\multiput(558.00,680.17)(14.440,6.000){2}{\rule{0.617pt}{0.400pt}}
\multiput(575.00,687.59)(1.395,0.482){9}{\rule{1.167pt}{0.116pt}}
\multiput(575.00,686.17)(13.579,6.000){2}{\rule{0.583pt}{0.400pt}}
\multiput(591.00,693.59)(1.255,0.485){11}{\rule{1.071pt}{0.117pt}}
\multiput(591.00,692.17)(14.776,7.000){2}{\rule{0.536pt}{0.400pt}}
\multiput(608.00,700.59)(1.179,0.485){11}{\rule{1.014pt}{0.117pt}}
\multiput(608.00,699.17)(13.895,7.000){2}{\rule{0.507pt}{0.400pt}}
\multiput(624.00,707.59)(1.255,0.485){11}{\rule{1.071pt}{0.117pt}}
\multiput(624.00,706.17)(14.776,7.000){2}{\rule{0.536pt}{0.400pt}}
\multiput(641.00,714.59)(1.022,0.488){13}{\rule{0.900pt}{0.117pt}}
\multiput(641.00,713.17)(14.132,8.000){2}{\rule{0.450pt}{0.400pt}}
\multiput(657.00,722.59)(1.255,0.485){11}{\rule{1.071pt}{0.117pt}}
\multiput(657.00,721.17)(14.776,7.000){2}{\rule{0.536pt}{0.400pt}}
\multiput(674.00,729.59)(1.179,0.485){11}{\rule{1.014pt}{0.117pt}}
\multiput(674.00,728.17)(13.895,7.000){2}{\rule{0.507pt}{0.400pt}}
\multiput(690.00,736.59)(1.485,0.482){9}{\rule{1.233pt}{0.116pt}}
\multiput(690.00,735.17)(14.440,6.000){2}{\rule{0.617pt}{0.400pt}}
\multiput(707.00,742.59)(1.395,0.482){9}{\rule{1.167pt}{0.116pt}}
\multiput(707.00,741.17)(13.579,6.000){2}{\rule{0.583pt}{0.400pt}}
\multiput(723.00,748.60)(2.382,0.468){5}{\rule{1.800pt}{0.113pt}}
\multiput(723.00,747.17)(13.264,4.000){2}{\rule{0.900pt}{0.400pt}}
\multiput(740.00,752.60)(2.382,0.468){5}{\rule{1.800pt}{0.113pt}}
\multiput(740.00,751.17)(13.264,4.000){2}{\rule{0.900pt}{0.400pt}}
\multiput(757.00,756.61)(3.365,0.447){3}{\rule{2.233pt}{0.108pt}}
\multiput(757.00,755.17)(11.365,3.000){2}{\rule{1.117pt}{0.400pt}}
\put(773,759.17){\rule{3.500pt}{0.400pt}}
\multiput(773.00,758.17)(9.736,2.000){2}{\rule{1.750pt}{0.400pt}}
\put(790,760.67){\rule{3.854pt}{0.400pt}}
\multiput(790.00,760.17)(8.000,1.000){2}{\rule{1.927pt}{0.400pt}}
\put(806,761.67){\rule{4.095pt}{0.400pt}}
\multiput(806.00,761.17)(8.500,1.000){2}{\rule{2.048pt}{0.400pt}}
\put(823,762.67){\rule{3.854pt}{0.400pt}}
\multiput(823.00,762.17)(8.000,1.000){2}{\rule{1.927pt}{0.400pt}}
\put(839,763.67){\rule{4.095pt}{0.400pt}}
\multiput(839.00,763.17)(8.500,1.000){2}{\rule{2.048pt}{0.400pt}}
\put(326.0,652.0){\rule[-0.200pt]{4.095pt}{0.400pt}}
\put(872,764.67){\rule{4.095pt}{0.400pt}}
\multiput(872.00,764.17)(8.500,1.000){2}{\rule{2.048pt}{0.400pt}}
\put(856.0,765.0){\rule[-0.200pt]{3.854pt}{0.400pt}}
\put(939,765.67){\rule{3.854pt}{0.400pt}}
\multiput(939.00,765.17)(8.000,1.000){2}{\rule{1.927pt}{0.400pt}}
\put(889.0,766.0){\rule[-0.200pt]{12.045pt}{0.400pt}}
\put(1038,766.67){\rule{3.854pt}{0.400pt}}
\multiput(1038.00,766.17)(8.000,1.000){2}{\rule{1.927pt}{0.400pt}}
\put(955.0,767.0){\rule[-0.200pt]{19.995pt}{0.400pt}}
\put(1054.0,768.0){\rule[-0.200pt]{19.995pt}{0.400pt}}
\end{picture}
\vspace{0.5cm}
\caption{Illustrative plot for matter effects on $|V_{e1}|$
associated with neutrinos ($+A$) and antineutrinos ($-A$),
where $\Delta m^2_{21} = 5\cdot 10^{-5} ~ {\rm eV}^2$ and 
$\Delta m^2_{31} = 3\cdot 10^{-3} ~ {\rm eV}^2$ have been input.}
\end{figure}
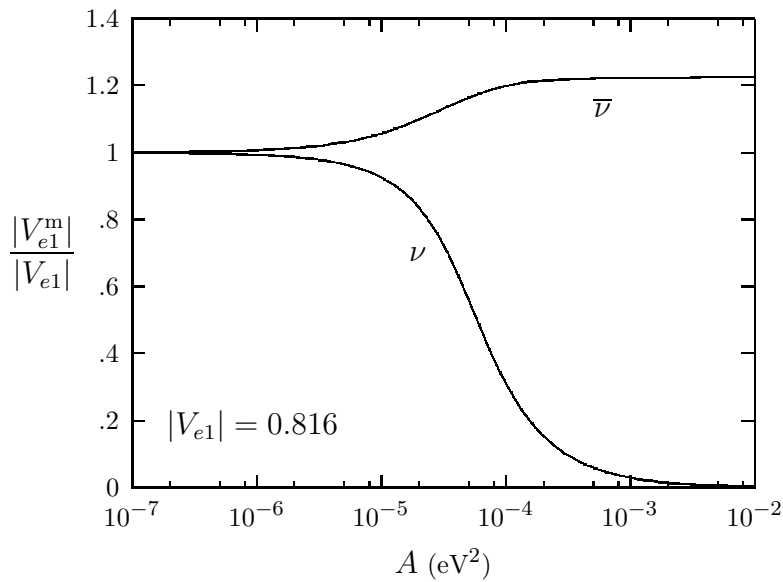
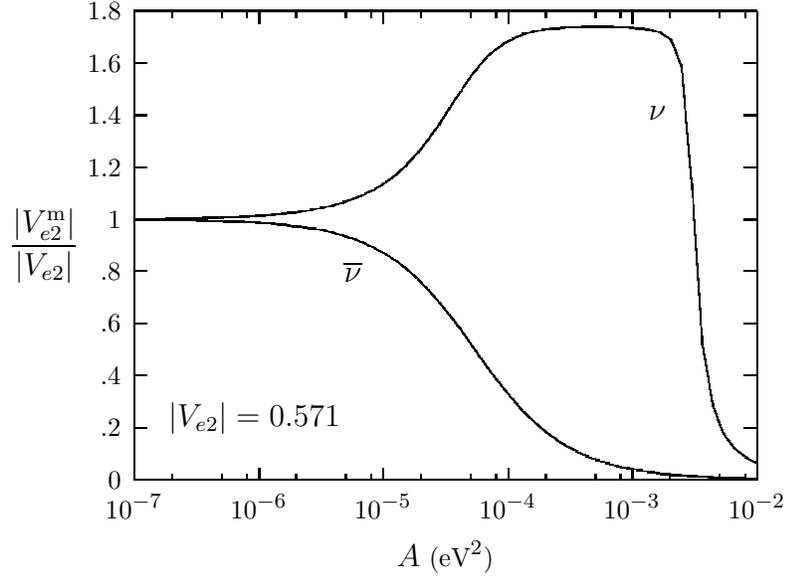
\begin{figure}
\setlength{\unitlength}{0.240900pt}
\ifx\plotpoint\undefined\newsavebox{\plotpoint}\fi
\sbox{\plotpoint}{\rule[-0.200pt]{0.400pt}{0.400pt}}%
\begin{picture}(1200,900)(-300,0)
\font\gnuplot=cmr10 at 10pt
\gnuplot
\sbox{\plotpoint}{\rule[-0.200pt]{0.400pt}{0.400pt}}%
\put(161.0,123.0){\rule[-0.200pt]{4.818pt}{0.400pt}}
\put(141,123){\makebox(0,0)[r]{0}}
\put(1119.0,123.0){\rule[-0.200pt]{4.818pt}{0.400pt}}
\put(161.0,205.0){\rule[-0.200pt]{4.818pt}{0.400pt}}
\put(141,205){\makebox(0,0)[r]{.2}}
\put(1119.0,205.0){\rule[-0.200pt]{4.818pt}{0.400pt}}
\put(161.0,287.0){\rule[-0.200pt]{4.818pt}{0.400pt}}
\put(141,287){\makebox(0,0)[r]{.4}}
\put(1119.0,287.0){\rule[-0.200pt]{4.818pt}{0.400pt}}
\put(161.0,369.0){\rule[-0.200pt]{4.818pt}{0.400pt}}
\put(141,369){\makebox(0,0)[r]{.6}}
\put(1119.0,369.0){\rule[-0.200pt]{4.818pt}{0.400pt}}
\put(161.0,451.0){\rule[-0.200pt]{4.818pt}{0.400pt}}
\put(141,451){\makebox(0,0)[r]{.8}}
\put(1119.0,451.0){\rule[-0.200pt]{4.818pt}{0.400pt}}
\put(161.0,532.0){\rule[-0.200pt]{4.818pt}{0.400pt}}
\put(141,532){\makebox(0,0)[r]{1}}
\put(1119.0,532.0){\rule[-0.200pt]{4.818pt}{0.400pt}}
\put(161.0,614.0){\rule[-0.200pt]{4.818pt}{0.400pt}}
\put(141,614){\makebox(0,0)[r]{1.2}}
\put(1119.0,614.0){\rule[-0.200pt]{4.818pt}{0.400pt}}
\put(161.0,696.0){\rule[-0.200pt]{4.818pt}{0.400pt}}
\put(141,696){\makebox(0,0)[r]{1.4}}
\put(1119.0,696.0){\rule[-0.200pt]{4.818pt}{0.400pt}}
\put(161.0,778.0){\rule[-0.200pt]{4.818pt}{0.400pt}}
\put(141,778){\makebox(0,0)[r]{1.6}}
\put(1119.0,778.0){\rule[-0.200pt]{4.818pt}{0.400pt}}
\put(161.0,860.0){\rule[-0.200pt]{4.818pt}{0.400pt}}
\put(141,860){\makebox(0,0)[r]{1.8}}
\put(1119.0,860.0){\rule[-0.200pt]{4.818pt}{0.400pt}}
\put(161.0,123.0){\rule[-0.200pt]{0.400pt}{4.818pt}}
\put(161,82){\makebox(0,0){10$^{-7}$}}
\put(161.0,840.0){\rule[-0.200pt]{0.400pt}{4.818pt}}
\put(220.0,123.0){\rule[-0.200pt]{0.400pt}{2.409pt}}
\put(220.0,850.0){\rule[-0.200pt]{0.400pt}{2.409pt}}
\put(298.0,123.0){\rule[-0.200pt]{0.400pt}{2.409pt}}
\put(298.0,850.0){\rule[-0.200pt]{0.400pt}{2.409pt}}
\put(338.0,123.0){\rule[-0.200pt]{0.400pt}{2.409pt}}
\put(338.0,850.0){\rule[-0.200pt]{0.400pt}{2.409pt}}
\put(357.0,123.0){\rule[-0.200pt]{0.400pt}{4.818pt}}
\put(357,82){\makebox(0,0){10$^{-6}$}}
\put(357.0,840.0){\rule[-0.200pt]{0.400pt}{4.818pt}}
\put(415.0,123.0){\rule[-0.200pt]{0.400pt}{2.409pt}}
\put(415.0,850.0){\rule[-0.200pt]{0.400pt}{2.409pt}}
\put(493.0,123.0){\rule[-0.200pt]{0.400pt}{2.409pt}}
\put(493.0,850.0){\rule[-0.200pt]{0.400pt}{2.409pt}}
\put(533.0,123.0){\rule[-0.200pt]{0.400pt}{2.409pt}}
\put(533.0,850.0){\rule[-0.200pt]{0.400pt}{2.409pt}}
\put(552.0,123.0){\rule[-0.200pt]{0.400pt}{4.818pt}}
\put(552,82){\makebox(0,0){10$^{-5}$}}
\put(552.0,840.0){\rule[-0.200pt]{0.400pt}{4.818pt}}
\put(611.0,123.0){\rule[-0.200pt]{0.400pt}{2.409pt}}
\put(611.0,850.0){\rule[-0.200pt]{0.400pt}{2.409pt}}
\put(689.0,123.0){\rule[-0.200pt]{0.400pt}{2.409pt}}
\put(689.0,850.0){\rule[-0.200pt]{0.400pt}{2.409pt}}
\put(729.0,123.0){\rule[-0.200pt]{0.400pt}{2.409pt}}
\put(729.0,850.0){\rule[-0.200pt]{0.400pt}{2.409pt}}
\put(748.0,123.0){\rule[-0.200pt]{0.400pt}{4.818pt}}
\put(748,82){\makebox(0,0){10$^{-4}$}}
\put(748.0,840.0){\rule[-0.200pt]{0.400pt}{4.818pt}}
\put(807.0,123.0){\rule[-0.200pt]{0.400pt}{2.409pt}}
\put(807.0,850.0){\rule[-0.200pt]{0.400pt}{2.409pt}}
\put(885.0,123.0){\rule[-0.200pt]{0.400pt}{2.409pt}}
\put(885.0,850.0){\rule[-0.200pt]{0.400pt}{2.409pt}}
\put(924.0,123.0){\rule[-0.200pt]{0.400pt}{2.409pt}}
\put(924.0,850.0){\rule[-0.200pt]{0.400pt}{2.409pt}}
\put(943.0,123.0){\rule[-0.200pt]{0.400pt}{4.818pt}}
\put(943,82){\makebox(0,0){10$^{-3}$}}
\put(943.0,840.0){\rule[-0.200pt]{0.400pt}{4.818pt}}
\put(1002.0,123.0){\rule[-0.200pt]{0.400pt}{2.409pt}}
\put(1002.0,850.0){\rule[-0.200pt]{0.400pt}{2.409pt}}
\put(1080.0,123.0){\rule[-0.200pt]{0.400pt}{2.409pt}}
\put(1080.0,850.0){\rule[-0.200pt]{0.400pt}{2.409pt}}
\put(1120.0,123.0){\rule[-0.200pt]{0.400pt}{2.409pt}}
\put(1120.0,850.0){\rule[-0.200pt]{0.400pt}{2.409pt}}
\put(1139.0,123.0){\rule[-0.200pt]{0.400pt}{4.818pt}}
\put(1139,82){\makebox(0,0){10$^{-2}$}}
\put(1139.0,840.0){\rule[-0.200pt]{0.400pt}{4.818pt}}
\put(161.0,123.0){\rule[-0.200pt]{235.600pt}{0.400pt}}
\put(1139.0,123.0){\rule[-0.200pt]{0.400pt}{177.543pt}}
\put(161.0,860.0){\rule[-0.200pt]{235.600pt}{0.400pt}}
\put(20,491){\makebox(0,0){$\displaystyle\frac{|V^{\rm m}_{e2}|}{|V_{e2}|}$}}
\put(650,3){\makebox(0,0){$A$ (eV$^2$)}}
\put(505,450){\makebox(0,0){$\overline{\nu}$}}
\put(983,700){\makebox(0,0){$\nu$}}
\put(350,220){\makebox(0,0){$|V_{e2}| = 0.571$}}
\put(161.0,123.0){\rule[-0.200pt]{0.400pt}{177.543pt}}
\put(161,533){\usebox{\plotpoint}}
\put(211,532.67){\rule{3.854pt}{0.400pt}}
\multiput(211.00,532.17)(8.000,1.000){2}{\rule{1.927pt}{0.400pt}}
\put(161.0,533.0){\rule[-0.200pt]{12.045pt}{0.400pt}}
\put(260,533.67){\rule{4.095pt}{0.400pt}}
\multiput(260.00,533.17)(8.500,1.000){2}{\rule{2.048pt}{0.400pt}}
\put(227.0,534.0){\rule[-0.200pt]{7.950pt}{0.400pt}}
\put(293,534.67){\rule{4.095pt}{0.400pt}}
\multiput(293.00,534.17)(8.500,1.000){2}{\rule{2.048pt}{0.400pt}}
\put(277.0,535.0){\rule[-0.200pt]{3.854pt}{0.400pt}}
\put(326,535.67){\rule{4.095pt}{0.400pt}}
\multiput(326.00,535.17)(8.500,1.000){2}{\rule{2.048pt}{0.400pt}}
\put(343,536.67){\rule{4.095pt}{0.400pt}}
\multiput(343.00,536.17)(8.500,1.000){2}{\rule{2.048pt}{0.400pt}}
\put(360,537.67){\rule{3.854pt}{0.400pt}}
\multiput(360.00,537.17)(8.000,1.000){2}{\rule{1.927pt}{0.400pt}}
\put(376,539.17){\rule{3.500pt}{0.400pt}}
\multiput(376.00,538.17)(9.736,2.000){2}{\rule{1.750pt}{0.400pt}}
\put(393,541.17){\rule{3.300pt}{0.400pt}}
\multiput(393.00,540.17)(9.151,2.000){2}{\rule{1.650pt}{0.400pt}}
\put(409,543.17){\rule{3.500pt}{0.400pt}}
\multiput(409.00,542.17)(9.736,2.000){2}{\rule{1.750pt}{0.400pt}}
\multiput(426.00,545.61)(3.365,0.447){3}{\rule{2.233pt}{0.108pt}}
\multiput(426.00,544.17)(11.365,3.000){2}{\rule{1.117pt}{0.400pt}}
\multiput(442.00,548.61)(3.588,0.447){3}{\rule{2.367pt}{0.108pt}}
\multiput(442.00,547.17)(12.088,3.000){2}{\rule{1.183pt}{0.400pt}}
\multiput(459.00,551.60)(2.236,0.468){5}{\rule{1.700pt}{0.113pt}}
\multiput(459.00,550.17)(12.472,4.000){2}{\rule{0.850pt}{0.400pt}}
\multiput(475.00,555.59)(1.823,0.477){7}{\rule{1.460pt}{0.115pt}}
\multiput(475.00,554.17)(13.970,5.000){2}{\rule{0.730pt}{0.400pt}}
\multiput(492.00,560.59)(1.395,0.482){9}{\rule{1.167pt}{0.116pt}}
\multiput(492.00,559.17)(13.579,6.000){2}{\rule{0.583pt}{0.400pt}}
\multiput(508.00,566.59)(1.255,0.485){11}{\rule{1.071pt}{0.117pt}}
\multiput(508.00,565.17)(14.776,7.000){2}{\rule{0.536pt}{0.400pt}}
\multiput(525.00,573.59)(1.022,0.488){13}{\rule{0.900pt}{0.117pt}}
\multiput(525.00,572.17)(14.132,8.000){2}{\rule{0.450pt}{0.400pt}}
\multiput(541.00,581.58)(0.779,0.492){19}{\rule{0.718pt}{0.118pt}}
\multiput(541.00,580.17)(15.509,11.000){2}{\rule{0.359pt}{0.400pt}}
\multiput(558.00,592.58)(0.655,0.493){23}{\rule{0.623pt}{0.119pt}}
\multiput(558.00,591.17)(15.707,13.000){2}{\rule{0.312pt}{0.400pt}}
\multiput(575.00,605.58)(0.531,0.494){27}{\rule{0.527pt}{0.119pt}}
\multiput(575.00,604.17)(14.907,15.000){2}{\rule{0.263pt}{0.400pt}}
\multiput(591.58,620.00)(0.495,0.528){31}{\rule{0.119pt}{0.524pt}}
\multiput(590.17,620.00)(17.000,16.913){2}{\rule{0.400pt}{0.262pt}}
\multiput(608.58,638.00)(0.494,0.657){29}{\rule{0.119pt}{0.625pt}}
\multiput(607.17,638.00)(16.000,19.703){2}{\rule{0.400pt}{0.313pt}}
\multiput(624.58,659.00)(0.495,0.678){31}{\rule{0.119pt}{0.641pt}}
\multiput(623.17,659.00)(17.000,21.669){2}{\rule{0.400pt}{0.321pt}}
\multiput(641.58,682.00)(0.494,0.817){29}{\rule{0.119pt}{0.750pt}}
\multiput(640.17,682.00)(16.000,24.443){2}{\rule{0.400pt}{0.375pt}}
\multiput(657.58,708.00)(0.495,0.738){31}{\rule{0.119pt}{0.688pt}}
\multiput(656.17,708.00)(17.000,23.572){2}{\rule{0.400pt}{0.344pt}}
\multiput(674.58,733.00)(0.494,0.753){29}{\rule{0.119pt}{0.700pt}}
\multiput(673.17,733.00)(16.000,22.547){2}{\rule{0.400pt}{0.350pt}}
\multiput(690.58,757.00)(0.495,0.618){31}{\rule{0.119pt}{0.594pt}}
\multiput(689.17,757.00)(17.000,19.767){2}{\rule{0.400pt}{0.297pt}}
\multiput(707.58,778.00)(0.494,0.529){29}{\rule{0.119pt}{0.525pt}}
\multiput(706.17,778.00)(16.000,15.910){2}{\rule{0.400pt}{0.263pt}}
\multiput(723.00,795.58)(0.655,0.493){23}{\rule{0.623pt}{0.119pt}}
\multiput(723.00,794.17)(15.707,13.000){2}{\rule{0.312pt}{0.400pt}}
\multiput(740.00,808.59)(0.961,0.489){15}{\rule{0.856pt}{0.118pt}}
\multiput(740.00,807.17)(15.224,9.000){2}{\rule{0.428pt}{0.400pt}}
\multiput(757.00,817.59)(1.179,0.485){11}{\rule{1.014pt}{0.117pt}}
\multiput(757.00,816.17)(13.895,7.000){2}{\rule{0.507pt}{0.400pt}}
\multiput(773.00,824.60)(2.382,0.468){5}{\rule{1.800pt}{0.113pt}}
\multiput(773.00,823.17)(13.264,4.000){2}{\rule{0.900pt}{0.400pt}}
\multiput(790.00,828.61)(3.365,0.447){3}{\rule{2.233pt}{0.108pt}}
\multiput(790.00,827.17)(11.365,3.000){2}{\rule{1.117pt}{0.400pt}}
\put(806,830.67){\rule{4.095pt}{0.400pt}}
\multiput(806.00,830.17)(8.500,1.000){2}{\rule{2.048pt}{0.400pt}}
\put(823,832.17){\rule{3.300pt}{0.400pt}}
\multiput(823.00,831.17)(9.151,2.000){2}{\rule{1.650pt}{0.400pt}}
\put(310.0,536.0){\rule[-0.200pt]{3.854pt}{0.400pt}}
\put(856,833.67){\rule{3.854pt}{0.400pt}}
\multiput(856.00,833.17)(8.000,1.000){2}{\rule{1.927pt}{0.400pt}}
\put(839.0,834.0){\rule[-0.200pt]{4.095pt}{0.400pt}}
\put(905,833.67){\rule{4.095pt}{0.400pt}}
\multiput(905.00,834.17)(8.500,-1.000){2}{\rule{2.048pt}{0.400pt}}
\put(872.0,835.0){\rule[-0.200pt]{7.950pt}{0.400pt}}
\put(939,832.17){\rule{3.300pt}{0.400pt}}
\multiput(939.00,833.17)(9.151,-2.000){2}{\rule{1.650pt}{0.400pt}}
\put(955,830.17){\rule{3.500pt}{0.400pt}}
\multiput(955.00,831.17)(9.736,-2.000){2}{\rule{1.750pt}{0.400pt}}
\multiput(972.00,828.94)(2.236,-0.468){5}{\rule{1.700pt}{0.113pt}}
\multiput(972.00,829.17)(12.472,-4.000){2}{\rule{0.850pt}{0.400pt}}
\multiput(988.00,824.92)(0.712,-0.492){21}{\rule{0.667pt}{0.119pt}}
\multiput(988.00,825.17)(15.616,-12.000){2}{\rule{0.333pt}{0.400pt}}
\multiput(1005.58,809.33)(0.494,-1.298){29}{\rule{0.119pt}{1.125pt}}
\multiput(1004.17,811.67)(16.000,-38.665){2}{\rule{0.400pt}{0.563pt}}
\multiput(1021.58,753.64)(0.495,-5.821){31}{\rule{0.119pt}{4.665pt}}
\multiput(1020.17,763.32)(17.000,-184.318){2}{\rule{0.400pt}{2.332pt}}
\multiput(1038.58,553.37)(0.494,-7.763){29}{\rule{0.119pt}{6.175pt}}
\multiput(1037.17,566.18)(16.000,-230.183){2}{\rule{0.400pt}{3.088pt}}
\multiput(1054.58,326.01)(0.495,-2.934){31}{\rule{0.119pt}{2.406pt}}
\multiput(1053.17,331.01)(17.000,-93.006){2}{\rule{0.400pt}{1.203pt}}
\multiput(1071.58,233.23)(0.494,-1.330){29}{\rule{0.119pt}{1.150pt}}
\multiput(1070.17,235.61)(16.000,-39.613){2}{\rule{0.400pt}{0.575pt}}
\multiput(1087.58,193.34)(0.495,-0.678){31}{\rule{0.119pt}{0.641pt}}
\multiput(1086.17,194.67)(17.000,-21.669){2}{\rule{0.400pt}{0.321pt}}
\multiput(1104.00,171.92)(0.607,-0.494){25}{\rule{0.586pt}{0.119pt}}
\multiput(1104.00,172.17)(15.784,-14.000){2}{\rule{0.293pt}{0.400pt}}
\multiput(1121.00,157.93)(0.902,-0.489){15}{\rule{0.811pt}{0.118pt}}
\multiput(1121.00,158.17)(14.316,-9.000){2}{\rule{0.406pt}{0.400pt}}
\put(922.0,834.0){\rule[-0.200pt]{4.095pt}{0.400pt}}
\put(161,532){\usebox{\plotpoint}}
\put(194,530.67){\rule{4.095pt}{0.400pt}}
\multiput(194.00,531.17)(8.500,-1.000){2}{\rule{2.048pt}{0.400pt}}
\put(161.0,532.0){\rule[-0.200pt]{7.950pt}{0.400pt}}
\put(260,529.67){\rule{4.095pt}{0.400pt}}
\multiput(260.00,530.17)(8.500,-1.000){2}{\rule{2.048pt}{0.400pt}}
\put(211.0,531.0){\rule[-0.200pt]{11.804pt}{0.400pt}}
\put(293,528.67){\rule{4.095pt}{0.400pt}}
\multiput(293.00,529.17)(8.500,-1.000){2}{\rule{2.048pt}{0.400pt}}
\put(277.0,530.0){\rule[-0.200pt]{3.854pt}{0.400pt}}
\put(326,527.67){\rule{4.095pt}{0.400pt}}
\multiput(326.00,528.17)(8.500,-1.000){2}{\rule{2.048pt}{0.400pt}}
\put(343,526.67){\rule{4.095pt}{0.400pt}}
\multiput(343.00,527.17)(8.500,-1.000){2}{\rule{2.048pt}{0.400pt}}
\put(360,525.67){\rule{3.854pt}{0.400pt}}
\multiput(360.00,526.17)(8.000,-1.000){2}{\rule{1.927pt}{0.400pt}}
\put(376,524.17){\rule{3.500pt}{0.400pt}}
\multiput(376.00,525.17)(9.736,-2.000){2}{\rule{1.750pt}{0.400pt}}
\put(393,522.17){\rule{3.300pt}{0.400pt}}
\multiput(393.00,523.17)(9.151,-2.000){2}{\rule{1.650pt}{0.400pt}}
\put(409,520.17){\rule{3.500pt}{0.400pt}}
\multiput(409.00,521.17)(9.736,-2.000){2}{\rule{1.750pt}{0.400pt}}
\put(426,518.17){\rule{3.300pt}{0.400pt}}
\multiput(426.00,519.17)(9.151,-2.000){2}{\rule{1.650pt}{0.400pt}}
\multiput(442.00,516.95)(3.588,-0.447){3}{\rule{2.367pt}{0.108pt}}
\multiput(442.00,517.17)(12.088,-3.000){2}{\rule{1.183pt}{0.400pt}}
\multiput(459.00,513.94)(2.236,-0.468){5}{\rule{1.700pt}{0.113pt}}
\multiput(459.00,514.17)(12.472,-4.000){2}{\rule{0.850pt}{0.400pt}}
\multiput(475.00,509.93)(1.823,-0.477){7}{\rule{1.460pt}{0.115pt}}
\multiput(475.00,510.17)(13.970,-5.000){2}{\rule{0.730pt}{0.400pt}}
\multiput(492.00,504.93)(1.712,-0.477){7}{\rule{1.380pt}{0.115pt}}
\multiput(492.00,505.17)(13.136,-5.000){2}{\rule{0.690pt}{0.400pt}}
\multiput(508.00,499.93)(1.255,-0.485){11}{\rule{1.071pt}{0.117pt}}
\multiput(508.00,500.17)(14.776,-7.000){2}{\rule{0.536pt}{0.400pt}}
\multiput(525.00,492.93)(1.022,-0.488){13}{\rule{0.900pt}{0.117pt}}
\multiput(525.00,493.17)(14.132,-8.000){2}{\rule{0.450pt}{0.400pt}}
\multiput(541.00,484.93)(0.961,-0.489){15}{\rule{0.856pt}{0.118pt}}
\multiput(541.00,485.17)(15.224,-9.000){2}{\rule{0.428pt}{0.400pt}}
\multiput(558.00,475.92)(0.779,-0.492){19}{\rule{0.718pt}{0.118pt}}
\multiput(558.00,476.17)(15.509,-11.000){2}{\rule{0.359pt}{0.400pt}}
\multiput(575.00,464.92)(0.616,-0.493){23}{\rule{0.592pt}{0.119pt}}
\multiput(575.00,465.17)(14.771,-13.000){2}{\rule{0.296pt}{0.400pt}}
\multiput(591.00,451.92)(0.566,-0.494){27}{\rule{0.553pt}{0.119pt}}
\multiput(591.00,452.17)(15.852,-15.000){2}{\rule{0.277pt}{0.400pt}}
\multiput(608.58,435.82)(0.494,-0.529){29}{\rule{0.119pt}{0.525pt}}
\multiput(607.17,436.91)(16.000,-15.910){2}{\rule{0.400pt}{0.263pt}}
\multiput(624.58,418.73)(0.495,-0.558){31}{\rule{0.119pt}{0.547pt}}
\multiput(623.17,419.86)(17.000,-17.865){2}{\rule{0.400pt}{0.274pt}}
\multiput(641.58,399.51)(0.494,-0.625){29}{\rule{0.119pt}{0.600pt}}
\multiput(640.17,400.75)(16.000,-18.755){2}{\rule{0.400pt}{0.300pt}}
\multiput(657.58,379.44)(0.495,-0.648){31}{\rule{0.119pt}{0.618pt}}
\multiput(656.17,380.72)(17.000,-20.718){2}{\rule{0.400pt}{0.309pt}}
\multiput(674.58,357.20)(0.494,-0.721){29}{\rule{0.119pt}{0.675pt}}
\multiput(673.17,358.60)(16.000,-21.599){2}{\rule{0.400pt}{0.338pt}}
\multiput(690.58,334.24)(0.495,-0.708){31}{\rule{0.119pt}{0.665pt}}
\multiput(689.17,335.62)(17.000,-22.620){2}{\rule{0.400pt}{0.332pt}}
\multiput(707.58,310.30)(0.494,-0.689){29}{\rule{0.119pt}{0.650pt}}
\multiput(706.17,311.65)(16.000,-20.651){2}{\rule{0.400pt}{0.325pt}}
\multiput(723.58,288.44)(0.495,-0.648){31}{\rule{0.119pt}{0.618pt}}
\multiput(722.17,289.72)(17.000,-20.718){2}{\rule{0.400pt}{0.309pt}}
\multiput(740.58,266.63)(0.495,-0.588){31}{\rule{0.119pt}{0.571pt}}
\multiput(739.17,267.82)(17.000,-18.816){2}{\rule{0.400pt}{0.285pt}}
\multiput(757.58,246.61)(0.494,-0.593){29}{\rule{0.119pt}{0.575pt}}
\multiput(756.17,247.81)(16.000,-17.807){2}{\rule{0.400pt}{0.288pt}}
\multiput(773.00,228.92)(0.529,-0.494){29}{\rule{0.525pt}{0.119pt}}
\multiput(773.00,229.17)(15.910,-16.000){2}{\rule{0.263pt}{0.400pt}}
\multiput(790.00,212.92)(0.570,-0.494){25}{\rule{0.557pt}{0.119pt}}
\multiput(790.00,213.17)(14.844,-14.000){2}{\rule{0.279pt}{0.400pt}}
\multiput(806.00,198.92)(0.655,-0.493){23}{\rule{0.623pt}{0.119pt}}
\multiput(806.00,199.17)(15.707,-13.000){2}{\rule{0.312pt}{0.400pt}}
\multiput(823.00,185.92)(0.808,-0.491){17}{\rule{0.740pt}{0.118pt}}
\multiput(823.00,186.17)(14.464,-10.000){2}{\rule{0.370pt}{0.400pt}}
\multiput(839.00,175.93)(0.961,-0.489){15}{\rule{0.856pt}{0.118pt}}
\multiput(839.00,176.17)(15.224,-9.000){2}{\rule{0.428pt}{0.400pt}}
\multiput(856.00,166.93)(1.022,-0.488){13}{\rule{0.900pt}{0.117pt}}
\multiput(856.00,167.17)(14.132,-8.000){2}{\rule{0.450pt}{0.400pt}}
\multiput(872.00,158.93)(1.485,-0.482){9}{\rule{1.233pt}{0.116pt}}
\multiput(872.00,159.17)(14.440,-6.000){2}{\rule{0.617pt}{0.400pt}}
\multiput(889.00,152.93)(1.712,-0.477){7}{\rule{1.380pt}{0.115pt}}
\multiput(889.00,153.17)(13.136,-5.000){2}{\rule{0.690pt}{0.400pt}}
\multiput(905.00,147.93)(1.823,-0.477){7}{\rule{1.460pt}{0.115pt}}
\multiput(905.00,148.17)(13.970,-5.000){2}{\rule{0.730pt}{0.400pt}}
\multiput(922.00,142.95)(3.588,-0.447){3}{\rule{2.367pt}{0.108pt}}
\multiput(922.00,143.17)(12.088,-3.000){2}{\rule{1.183pt}{0.400pt}}
\multiput(939.00,139.95)(3.365,-0.447){3}{\rule{2.233pt}{0.108pt}}
\multiput(939.00,140.17)(11.365,-3.000){2}{\rule{1.117pt}{0.400pt}}
\multiput(955.00,136.95)(3.588,-0.447){3}{\rule{2.367pt}{0.108pt}}
\multiput(955.00,137.17)(12.088,-3.000){2}{\rule{1.183pt}{0.400pt}}
\put(972,133.17){\rule{3.300pt}{0.400pt}}
\multiput(972.00,134.17)(9.151,-2.000){2}{\rule{1.650pt}{0.400pt}}
\put(988,131.17){\rule{3.500pt}{0.400pt}}
\multiput(988.00,132.17)(9.736,-2.000){2}{\rule{1.750pt}{0.400pt}}
\put(1005,129.67){\rule{3.854pt}{0.400pt}}
\multiput(1005.00,130.17)(8.000,-1.000){2}{\rule{1.927pt}{0.400pt}}
\put(1021,128.67){\rule{4.095pt}{0.400pt}}
\multiput(1021.00,129.17)(8.500,-1.000){2}{\rule{2.048pt}{0.400pt}}
\put(1038,127.67){\rule{3.854pt}{0.400pt}}
\multiput(1038.00,128.17)(8.000,-1.000){2}{\rule{1.927pt}{0.400pt}}
\put(1054,126.67){\rule{4.095pt}{0.400pt}}
\multiput(1054.00,127.17)(8.500,-1.000){2}{\rule{2.048pt}{0.400pt}}
\put(1071,125.67){\rule{3.854pt}{0.400pt}}
\multiput(1071.00,126.17)(8.000,-1.000){2}{\rule{1.927pt}{0.400pt}}
\put(310.0,529.0){\rule[-0.200pt]{3.854pt}{0.400pt}}
\put(1104,124.67){\rule{4.095pt}{0.400pt}}
\multiput(1104.00,125.17)(8.500,-1.000){2}{\rule{2.048pt}{0.400pt}}
\put(1087.0,126.0){\rule[-0.200pt]{4.095pt}{0.400pt}}
\put(1121.0,125.0){\rule[-0.200pt]{3.854pt}{0.400pt}}
\end{picture}
\vspace{0.5cm}
\caption{Illustrative plot for matter effects on $|V_{e2}|$
associated with neutrinos ($+A$) and antineutrinos ($-A$),
where $\Delta m^2_{21} = 5\cdot 10^{-5} ~ {\rm eV}^2$ and 
$\Delta m^2_{31} = 3\cdot 10^{-3} ~ {\rm eV}^2$ have been input.}
\end{figure}
\begin{figure}
\setlength{\unitlength}{0.240900pt}
\ifx\plotpoint\undefined\newsavebox{\plotpoint}\fi
\sbox{\plotpoint}{\rule[-0.200pt]{0.400pt}{0.400pt}}%
\begin{picture}(1200,900)(-300,0)
\font\gnuplot=cmr10 at 10pt
\gnuplot
\sbox{\plotpoint}{\rule[-0.200pt]{0.400pt}{0.400pt}}%
\put(161.0,123.0){\rule[-0.200pt]{4.818pt}{0.400pt}}
\put(141,123){\makebox(0,0)[r]{0}}
\put(1119.0,123.0){\rule[-0.200pt]{4.818pt}{0.400pt}}
\put(161.0,190.0){\rule[-0.200pt]{4.818pt}{0.400pt}}
\put(141,190){\makebox(0,0)[r]{.1}}
\put(1119.0,190.0){\rule[-0.200pt]{4.818pt}{0.400pt}}
\put(161.0,257.0){\rule[-0.200pt]{4.818pt}{0.400pt}}
\put(141,257){\makebox(0,0)[r]{.2}}
\put(1119.0,257.0){\rule[-0.200pt]{4.818pt}{0.400pt}}
\put(161.0,324.0){\rule[-0.200pt]{4.818pt}{0.400pt}}
\put(141,324){\makebox(0,0)[r]{.3}}
\put(1119.0,324.0){\rule[-0.200pt]{4.818pt}{0.400pt}}
\put(161.0,391.0){\rule[-0.200pt]{4.818pt}{0.400pt}}
\put(141,391){\makebox(0,0)[r]{.4}}
\put(1119.0,391.0){\rule[-0.200pt]{4.818pt}{0.400pt}}
\put(161.0,458.0){\rule[-0.200pt]{4.818pt}{0.400pt}}
\put(141,458){\makebox(0,0)[r]{.5}}
\put(1119.0,458.0){\rule[-0.200pt]{4.818pt}{0.400pt}}
\put(161.0,525.0){\rule[-0.200pt]{4.818pt}{0.400pt}}
\put(141,525){\makebox(0,0)[r]{.6}}
\put(1119.0,525.0){\rule[-0.200pt]{4.818pt}{0.400pt}}
\put(161.0,592.0){\rule[-0.200pt]{4.818pt}{0.400pt}}
\put(141,592){\makebox(0,0)[r]{.7}}
\put(1119.0,592.0){\rule[-0.200pt]{4.818pt}{0.400pt}}
\put(161.0,659.0){\rule[-0.200pt]{4.818pt}{0.400pt}}
\put(141,659){\makebox(0,0)[r]{.8}}
\put(1119.0,659.0){\rule[-0.200pt]{4.818pt}{0.400pt}}
\put(161.0,726.0){\rule[-0.200pt]{4.818pt}{0.400pt}}
\put(141,726){\makebox(0,0)[r]{.9}}
\put(1119.0,726.0){\rule[-0.200pt]{4.818pt}{0.400pt}}
\put(161.0,793.0){\rule[-0.200pt]{4.818pt}{0.400pt}}
\put(141,793){\makebox(0,0)[r]{1}}
\put(1119.0,793.0){\rule[-0.200pt]{4.818pt}{0.400pt}}
\put(161.0,860.0){\rule[-0.200pt]{4.818pt}{0.400pt}}
\put(141,860){\makebox(0,0)[r]{1.1}}
\put(1119.0,860.0){\rule[-0.200pt]{4.818pt}{0.400pt}}
\put(161.0,123.0){\rule[-0.200pt]{0.400pt}{4.818pt}}
\put(161,82){\makebox(0,0){10$^{-7}$}}
\put(161.0,840.0){\rule[-0.200pt]{0.400pt}{4.818pt}}
\put(220.0,123.0){\rule[-0.200pt]{0.400pt}{2.409pt}}
\put(220.0,850.0){\rule[-0.200pt]{0.400pt}{2.409pt}}
\put(298.0,123.0){\rule[-0.200pt]{0.400pt}{2.409pt}}
\put(298.0,850.0){\rule[-0.200pt]{0.400pt}{2.409pt}}
\put(338.0,123.0){\rule[-0.200pt]{0.400pt}{2.409pt}}
\put(338.0,850.0){\rule[-0.200pt]{0.400pt}{2.409pt}}
\put(357.0,123.0){\rule[-0.200pt]{0.400pt}{4.818pt}}
\put(357,82){\makebox(0,0){10$^{-6}$}}
\put(357.0,840.0){\rule[-0.200pt]{0.400pt}{4.818pt}}
\put(415.0,123.0){\rule[-0.200pt]{0.400pt}{2.409pt}}
\put(415.0,850.0){\rule[-0.200pt]{0.400pt}{2.409pt}}
\put(493.0,123.0){\rule[-0.200pt]{0.400pt}{2.409pt}}
\put(493.0,850.0){\rule[-0.200pt]{0.400pt}{2.409pt}}
\put(533.0,123.0){\rule[-0.200pt]{0.400pt}{2.409pt}}
\put(533.0,850.0){\rule[-0.200pt]{0.400pt}{2.409pt}}
\put(552.0,123.0){\rule[-0.200pt]{0.400pt}{4.818pt}}
\put(552,82){\makebox(0,0){10$^{-5}$}}
\put(552.0,840.0){\rule[-0.200pt]{0.400pt}{4.818pt}}
\put(611.0,123.0){\rule[-0.200pt]{0.400pt}{2.409pt}}
\put(611.0,850.0){\rule[-0.200pt]{0.400pt}{2.409pt}}
\put(689.0,123.0){\rule[-0.200pt]{0.400pt}{2.409pt}}
\put(689.0,850.0){\rule[-0.200pt]{0.400pt}{2.409pt}}
\put(729.0,123.0){\rule[-0.200pt]{0.400pt}{2.409pt}}
\put(729.0,850.0){\rule[-0.200pt]{0.400pt}{2.409pt}}
\put(748.0,123.0){\rule[-0.200pt]{0.400pt}{4.818pt}}
\put(748,82){\makebox(0,0){10$^{-4}$}}
\put(748.0,840.0){\rule[-0.200pt]{0.400pt}{4.818pt}}
\put(807.0,123.0){\rule[-0.200pt]{0.400pt}{2.409pt}}
\put(807.0,850.0){\rule[-0.200pt]{0.400pt}{2.409pt}}
\put(885.0,123.0){\rule[-0.200pt]{0.400pt}{2.409pt}}
\put(885.0,850.0){\rule[-0.200pt]{0.400pt}{2.409pt}}
\put(924.0,123.0){\rule[-0.200pt]{0.400pt}{2.409pt}}
\put(924.0,850.0){\rule[-0.200pt]{0.400pt}{2.409pt}}
\put(943.0,123.0){\rule[-0.200pt]{0.400pt}{4.818pt}}
\put(943,82){\makebox(0,0){10$^{-3}$}}
\put(943.0,840.0){\rule[-0.200pt]{0.400pt}{4.818pt}}
\put(1002.0,123.0){\rule[-0.200pt]{0.400pt}{2.409pt}}
\put(1002.0,850.0){\rule[-0.200pt]{0.400pt}{2.409pt}}
\put(1080.0,123.0){\rule[-0.200pt]{0.400pt}{2.409pt}}
\put(1080.0,850.0){\rule[-0.200pt]{0.400pt}{2.409pt}}
\put(1120.0,123.0){\rule[-0.200pt]{0.400pt}{2.409pt}}
\put(1120.0,850.0){\rule[-0.200pt]{0.400pt}{2.409pt}}
\put(1139.0,123.0){\rule[-0.200pt]{0.400pt}{4.818pt}}
\put(1139,82){\makebox(0,0){10$^{-2}$}}
\put(1139.0,840.0){\rule[-0.200pt]{0.400pt}{4.818pt}}
\put(161.0,123.0){\rule[-0.200pt]{235.600pt}{0.400pt}}
\put(1139.0,123.0){\rule[-0.200pt]{0.400pt}{177.543pt}}
\put(161.0,860.0){\rule[-0.200pt]{235.600pt}{0.400pt}}
\put(0,491){\makebox(0,0){$\displaystyle\frac{|V^{\rm m}_{\mu 3}|}{|V_{\mu 3}|}$}}
\put(650,3){\makebox(0,0){$A$ (eV$^2$)}}
\put(1085,756){\makebox(0,0){$\overline{\nu}$}}
\put(998,500){\makebox(0,0){$\nu$}}
\put(350,220){\makebox(0,0){$|V_{\mu 3}| = 0.640$}}
\put(161.0,123.0){\rule[-0.200pt]{0.400pt}{177.543pt}}
\put(161,793){\usebox{\plotpoint}}
\put(823,791.67){\rule{3.854pt}{0.400pt}}
\multiput(823.00,792.17)(8.000,-1.000){2}{\rule{1.927pt}{0.400pt}}
\put(161.0,793.0){\rule[-0.200pt]{159.476pt}{0.400pt}}
\put(889,790.67){\rule{3.854pt}{0.400pt}}
\multiput(889.00,791.17)(8.000,-1.000){2}{\rule{1.927pt}{0.400pt}}
\put(839.0,792.0){\rule[-0.200pt]{12.045pt}{0.400pt}}
\put(922,789.67){\rule{4.095pt}{0.400pt}}
\multiput(922.00,790.17)(8.500,-1.000){2}{\rule{2.048pt}{0.400pt}}
\put(939,788.67){\rule{3.854pt}{0.400pt}}
\multiput(939.00,789.17)(8.000,-1.000){2}{\rule{1.927pt}{0.400pt}}
\put(955,787.17){\rule{3.500pt}{0.400pt}}
\multiput(955.00,788.17)(9.736,-2.000){2}{\rule{1.750pt}{0.400pt}}
\multiput(972.00,785.94)(2.236,-0.468){5}{\rule{1.700pt}{0.113pt}}
\multiput(972.00,786.17)(12.472,-4.000){2}{\rule{0.850pt}{0.400pt}}
\multiput(988.00,781.92)(0.779,-0.492){19}{\rule{0.718pt}{0.118pt}}
\multiput(988.00,782.17)(15.509,-11.000){2}{\rule{0.359pt}{0.400pt}}
\multiput(1005.58,767.54)(0.494,-1.233){29}{\rule{0.119pt}{1.075pt}}
\multiput(1004.17,769.77)(16.000,-36.769){2}{\rule{0.400pt}{0.538pt}}
\multiput(1021.58,714.81)(0.495,-5.460){31}{\rule{0.119pt}{4.382pt}}
\multiput(1020.17,723.90)(17.000,-172.904){2}{\rule{0.400pt}{2.191pt}}
\multiput(1038.58,526.92)(0.494,-7.283){29}{\rule{0.119pt}{5.800pt}}
\multiput(1037.17,538.96)(16.000,-215.962){2}{\rule{0.400pt}{2.900pt}}
\multiput(1054.58,313.60)(0.495,-2.753){31}{\rule{0.119pt}{2.265pt}}
\multiput(1053.17,318.30)(17.000,-87.299){2}{\rule{0.400pt}{1.132pt}}
\multiput(1071.58,226.43)(0.494,-1.265){29}{\rule{0.119pt}{1.100pt}}
\multiput(1070.17,228.72)(16.000,-37.717){2}{\rule{0.400pt}{0.550pt}}
\multiput(1087.58,188.53)(0.495,-0.618){31}{\rule{0.119pt}{0.594pt}}
\multiput(1086.17,189.77)(17.000,-19.767){2}{\rule{0.400pt}{0.297pt}}
\multiput(1104.00,168.92)(0.655,-0.493){23}{\rule{0.623pt}{0.119pt}}
\multiput(1104.00,169.17)(15.707,-13.000){2}{\rule{0.312pt}{0.400pt}}
\multiput(1121.00,155.93)(1.022,-0.488){13}{\rule{0.900pt}{0.117pt}}
\multiput(1121.00,156.17)(14.132,-8.000){2}{\rule{0.450pt}{0.400pt}}
\put(905.0,791.0){\rule[-0.200pt]{4.095pt}{0.400pt}}
\put(161,793){\usebox{\plotpoint}}
\put(839,792.67){\rule{4.095pt}{0.400pt}}
\multiput(839.00,792.17)(8.500,1.000){2}{\rule{2.048pt}{0.400pt}}
\put(161.0,793.0){\rule[-0.200pt]{163.330pt}{0.400pt}}
\put(972,793.67){\rule{3.854pt}{0.400pt}}
\multiput(972.00,793.17)(8.000,1.000){2}{\rule{1.927pt}{0.400pt}}
\put(856.0,794.0){\rule[-0.200pt]{27.944pt}{0.400pt}}
\put(988.0,795.0){\rule[-0.200pt]{35.894pt}{0.400pt}}
\end{picture}
\vspace{0.5cm}
\caption{Illustrative plot for matter effects on $|V_{\mu 3}|$
associated with neutrinos ($+A$) and antineutrinos ($-A$),
where $\Delta m^2_{21} = 5\cdot 10^{-5} ~ {\rm eV}^2$ and 
$\Delta m^2_{31} = 3\cdot 10^{-3} ~ {\rm eV}^2$ have been input.}
\end{figure}
\begin{figure}
\setlength{\unitlength}{0.240900pt}
\ifx\plotpoint\undefined\newsavebox{\plotpoint}\fi
\sbox{\plotpoint}{\rule[-0.200pt]{0.400pt}{0.400pt}}%
\begin{picture}(1200,900)(-300,0)
\font\gnuplot=cmr10 at 10pt
\gnuplot
\sbox{\plotpoint}{\rule[-0.200pt]{0.400pt}{0.400pt}}%
\put(161.0,123.0){\rule[-0.200pt]{4.818pt}{0.400pt}}
\put(141,123){\makebox(0,0)[r]{0}}
\put(1119.0,123.0){\rule[-0.200pt]{4.818pt}{0.400pt}}
\put(161.0,246.0){\rule[-0.200pt]{4.818pt}{0.400pt}}
\put(141,246){\makebox(0,0)[r]{.2}}
\put(1119.0,246.0){\rule[-0.200pt]{4.818pt}{0.400pt}}
\put(161.0,369.0){\rule[-0.200pt]{4.818pt}{0.400pt}}
\put(141,369){\makebox(0,0)[r]{.4}}
\put(1119.0,369.0){\rule[-0.200pt]{4.818pt}{0.400pt}}
\put(161.0,492.0){\rule[-0.200pt]{4.818pt}{0.400pt}}
\put(141,492){\makebox(0,0)[r]{.6}}
\put(1119.0,492.0){\rule[-0.200pt]{4.818pt}{0.400pt}}
\put(161.0,614.0){\rule[-0.200pt]{4.818pt}{0.400pt}}
\put(141,614){\makebox(0,0)[r]{.8}}
\put(1119.0,614.0){\rule[-0.200pt]{4.818pt}{0.400pt}}
\put(161.0,737.0){\rule[-0.200pt]{4.818pt}{0.400pt}}
\put(141,737){\makebox(0,0)[r]{1}}
\put(1119.0,737.0){\rule[-0.200pt]{4.818pt}{0.400pt}}
\put(161.0,860.0){\rule[-0.200pt]{4.818pt}{0.400pt}}
\put(141,860){\makebox(0,0)[r]{1.2}}
\put(1119.0,860.0){\rule[-0.200pt]{4.818pt}{0.400pt}}
\put(161.0,123.0){\rule[-0.200pt]{0.400pt}{4.818pt}}
\put(161,82){\makebox(0,0){10$^{-7}$}}
\put(161.0,840.0){\rule[-0.200pt]{0.400pt}{4.818pt}}
\put(220.0,123.0){\rule[-0.200pt]{0.400pt}{2.409pt}}
\put(220.0,850.0){\rule[-0.200pt]{0.400pt}{2.409pt}}
\put(298.0,123.0){\rule[-0.200pt]{0.400pt}{2.409pt}}
\put(298.0,850.0){\rule[-0.200pt]{0.400pt}{2.409pt}}
\put(338.0,123.0){\rule[-0.200pt]{0.400pt}{2.409pt}}
\put(338.0,850.0){\rule[-0.200pt]{0.400pt}{2.409pt}}
\put(357.0,123.0){\rule[-0.200pt]{0.400pt}{4.818pt}}
\put(357,82){\makebox(0,0){10$^{-6}$}}
\put(357.0,840.0){\rule[-0.200pt]{0.400pt}{4.818pt}}
\put(415.0,123.0){\rule[-0.200pt]{0.400pt}{2.409pt}}
\put(415.0,850.0){\rule[-0.200pt]{0.400pt}{2.409pt}}
\put(493.0,123.0){\rule[-0.200pt]{0.400pt}{2.409pt}}
\put(493.0,850.0){\rule[-0.200pt]{0.400pt}{2.409pt}}
\put(533.0,123.0){\rule[-0.200pt]{0.400pt}{2.409pt}}
\put(533.0,850.0){\rule[-0.200pt]{0.400pt}{2.409pt}}
\put(552.0,123.0){\rule[-0.200pt]{0.400pt}{4.818pt}}
\put(552,82){\makebox(0,0){10$^{-5}$}}
\put(552.0,840.0){\rule[-0.200pt]{0.400pt}{4.818pt}}
\put(611.0,123.0){\rule[-0.200pt]{0.400pt}{2.409pt}}
\put(611.0,850.0){\rule[-0.200pt]{0.400pt}{2.409pt}}
\put(689.0,123.0){\rule[-0.200pt]{0.400pt}{2.409pt}}
\put(689.0,850.0){\rule[-0.200pt]{0.400pt}{2.409pt}}
\put(729.0,123.0){\rule[-0.200pt]{0.400pt}{2.409pt}}
\put(729.0,850.0){\rule[-0.200pt]{0.400pt}{2.409pt}}
\put(748.0,123.0){\rule[-0.200pt]{0.400pt}{4.818pt}}
\put(748,82){\makebox(0,0){10$^{-4}$}}
\put(748.0,840.0){\rule[-0.200pt]{0.400pt}{4.818pt}}
\put(807.0,123.0){\rule[-0.200pt]{0.400pt}{2.409pt}}
\put(807.0,850.0){\rule[-0.200pt]{0.400pt}{2.409pt}}
\put(885.0,123.0){\rule[-0.200pt]{0.400pt}{2.409pt}}
\put(885.0,850.0){\rule[-0.200pt]{0.400pt}{2.409pt}}
\put(924.0,123.0){\rule[-0.200pt]{0.400pt}{2.409pt}}
\put(924.0,850.0){\rule[-0.200pt]{0.400pt}{2.409pt}}
\put(943.0,123.0){\rule[-0.200pt]{0.400pt}{4.818pt}}
\put(943,82){\makebox(0,0){10$^{-3}$}}
\put(943.0,840.0){\rule[-0.200pt]{0.400pt}{4.818pt}}
\put(1002.0,123.0){\rule[-0.200pt]{0.400pt}{2.409pt}}
\put(1002.0,850.0){\rule[-0.200pt]{0.400pt}{2.409pt}}
\put(1080.0,123.0){\rule[-0.200pt]{0.400pt}{2.409pt}}
\put(1080.0,850.0){\rule[-0.200pt]{0.400pt}{2.409pt}}
\put(1120.0,123.0){\rule[-0.200pt]{0.400pt}{2.409pt}}
\put(1120.0,850.0){\rule[-0.200pt]{0.400pt}{2.409pt}}
\put(1139.0,123.0){\rule[-0.200pt]{0.400pt}{4.818pt}}
\put(1139,82){\makebox(0,0){10$^{-2}$}}
\put(1139.0,840.0){\rule[-0.200pt]{0.400pt}{4.818pt}}
\put(161.0,123.0){\rule[-0.200pt]{235.600pt}{0.400pt}}
\put(1139.0,123.0){\rule[-0.200pt]{0.400pt}{177.543pt}}
\put(161.0,860.0){\rule[-0.200pt]{235.600pt}{0.400pt}}
\put(20,491){\makebox(0,0){$\displaystyle\frac{{\cal J}_{\rm m}}{\cal J}$}}
\put(650,3){\makebox(0,0){$A$ (eV$^2$)}}
\put(743,665){\makebox(0,0){$\nu$}}
\put(639,480){\makebox(0,0){$\overline{\nu}$}}
\put(350,220){\makebox(0,0){$|{\cal J}| = 0.020$}}
\put(161.0,123.0){\rule[-0.200pt]{0.400pt}{177.543pt}}
\put(161,738){\usebox{\plotpoint}}
\put(244,737.67){\rule{3.854pt}{0.400pt}}
\multiput(244.00,737.17)(8.000,1.000){2}{\rule{1.927pt}{0.400pt}}
\put(161.0,738.0){\rule[-0.200pt]{19.995pt}{0.400pt}}
\put(293,738.67){\rule{4.095pt}{0.400pt}}
\multiput(293.00,738.17)(8.500,1.000){2}{\rule{2.048pt}{0.400pt}}
\put(260.0,739.0){\rule[-0.200pt]{7.950pt}{0.400pt}}
\put(326,739.67){\rule{4.095pt}{0.400pt}}
\multiput(326.00,739.17)(8.500,1.000){2}{\rule{2.048pt}{0.400pt}}
\put(343,740.67){\rule{4.095pt}{0.400pt}}
\multiput(343.00,740.17)(8.500,1.000){2}{\rule{2.048pt}{0.400pt}}
\put(360,741.67){\rule{3.854pt}{0.400pt}}
\multiput(360.00,741.17)(8.000,1.000){2}{\rule{1.927pt}{0.400pt}}
\put(376,742.67){\rule{4.095pt}{0.400pt}}
\multiput(376.00,742.17)(8.500,1.000){2}{\rule{2.048pt}{0.400pt}}
\put(393,743.67){\rule{3.854pt}{0.400pt}}
\multiput(393.00,743.17)(8.000,1.000){2}{\rule{1.927pt}{0.400pt}}
\put(409,745.17){\rule{3.500pt}{0.400pt}}
\multiput(409.00,744.17)(9.736,2.000){2}{\rule{1.750pt}{0.400pt}}
\put(426,747.17){\rule{3.300pt}{0.400pt}}
\multiput(426.00,746.17)(9.151,2.000){2}{\rule{1.650pt}{0.400pt}}
\put(442,749.17){\rule{3.500pt}{0.400pt}}
\multiput(442.00,748.17)(9.736,2.000){2}{\rule{1.750pt}{0.400pt}}
\put(459,751.17){\rule{3.300pt}{0.400pt}}
\multiput(459.00,750.17)(9.151,2.000){2}{\rule{1.650pt}{0.400pt}}
\multiput(475.00,753.60)(2.382,0.468){5}{\rule{1.800pt}{0.113pt}}
\multiput(475.00,752.17)(13.264,4.000){2}{\rule{0.900pt}{0.400pt}}
\multiput(492.00,757.61)(3.365,0.447){3}{\rule{2.233pt}{0.108pt}}
\multiput(492.00,756.17)(11.365,3.000){2}{\rule{1.117pt}{0.400pt}}
\multiput(508.00,760.60)(2.382,0.468){5}{\rule{1.800pt}{0.113pt}}
\multiput(508.00,759.17)(13.264,4.000){2}{\rule{0.900pt}{0.400pt}}
\multiput(525.00,764.60)(2.236,0.468){5}{\rule{1.700pt}{0.113pt}}
\multiput(525.00,763.17)(12.472,4.000){2}{\rule{0.850pt}{0.400pt}}
\multiput(541.00,768.59)(1.823,0.477){7}{\rule{1.460pt}{0.115pt}}
\multiput(541.00,767.17)(13.970,5.000){2}{\rule{0.730pt}{0.400pt}}
\multiput(558.00,773.60)(2.382,0.468){5}{\rule{1.800pt}{0.113pt}}
\multiput(558.00,772.17)(13.264,4.000){2}{\rule{0.900pt}{0.400pt}}
\multiput(575.00,777.61)(3.365,0.447){3}{\rule{2.233pt}{0.108pt}}
\multiput(575.00,776.17)(11.365,3.000){2}{\rule{1.117pt}{0.400pt}}
\put(310.0,740.0){\rule[-0.200pt]{3.854pt}{0.400pt}}
\multiput(608.00,778.94)(2.236,-0.468){5}{\rule{1.700pt}{0.113pt}}
\multiput(608.00,779.17)(12.472,-4.000){2}{\rule{0.850pt}{0.400pt}}
\multiput(624.00,774.92)(0.779,-0.492){19}{\rule{0.718pt}{0.118pt}}
\multiput(624.00,775.17)(15.509,-11.000){2}{\rule{0.359pt}{0.400pt}}
\multiput(641.58,762.41)(0.494,-0.657){29}{\rule{0.119pt}{0.625pt}}
\multiput(640.17,763.70)(16.000,-19.703){2}{\rule{0.400pt}{0.313pt}}
\multiput(657.58,740.36)(0.495,-0.979){31}{\rule{0.119pt}{0.876pt}}
\multiput(656.17,742.18)(17.000,-31.181){2}{\rule{0.400pt}{0.438pt}}
\multiput(674.58,705.81)(0.494,-1.458){29}{\rule{0.119pt}{1.250pt}}
\multiput(673.17,708.41)(16.000,-43.406){2}{\rule{0.400pt}{0.625pt}}
\multiput(690.58,659.12)(0.495,-1.671){31}{\rule{0.119pt}{1.418pt}}
\multiput(689.17,662.06)(17.000,-53.058){2}{\rule{0.400pt}{0.709pt}}
\multiput(707.58,602.15)(0.494,-1.970){29}{\rule{0.119pt}{1.650pt}}
\multiput(706.17,605.58)(16.000,-58.575){2}{\rule{0.400pt}{0.825pt}}
\multiput(723.58,540.63)(0.495,-1.821){31}{\rule{0.119pt}{1.535pt}}
\multiput(722.17,543.81)(17.000,-57.813){2}{\rule{0.400pt}{0.768pt}}
\multiput(740.58,480.02)(0.495,-1.701){31}{\rule{0.119pt}{1.441pt}}
\multiput(739.17,483.01)(17.000,-54.009){2}{\rule{0.400pt}{0.721pt}}
\multiput(757.58,423.40)(0.494,-1.586){29}{\rule{0.119pt}{1.350pt}}
\multiput(756.17,426.20)(16.000,-47.198){2}{\rule{0.400pt}{0.675pt}}
\multiput(773.58,374.38)(0.495,-1.279){31}{\rule{0.119pt}{1.112pt}}
\multiput(772.17,376.69)(17.000,-40.692){2}{\rule{0.400pt}{0.556pt}}
\multiput(790.58,331.75)(0.494,-1.169){29}{\rule{0.119pt}{1.025pt}}
\multiput(789.17,333.87)(16.000,-34.873){2}{\rule{0.400pt}{0.513pt}}
\multiput(806.58,295.75)(0.495,-0.858){31}{\rule{0.119pt}{0.782pt}}
\multiput(805.17,297.38)(17.000,-27.376){2}{\rule{0.400pt}{0.391pt}}
\multiput(823.58,266.99)(0.494,-0.785){29}{\rule{0.119pt}{0.725pt}}
\multiput(822.17,268.50)(16.000,-23.495){2}{\rule{0.400pt}{0.363pt}}
\multiput(839.58,242.63)(0.495,-0.588){31}{\rule{0.119pt}{0.571pt}}
\multiput(838.17,243.82)(17.000,-18.816){2}{\rule{0.400pt}{0.285pt}}
\multiput(856.00,223.92)(0.497,-0.494){29}{\rule{0.500pt}{0.119pt}}
\multiput(856.00,224.17)(14.962,-16.000){2}{\rule{0.250pt}{0.400pt}}
\multiput(872.00,207.92)(0.655,-0.493){23}{\rule{0.623pt}{0.119pt}}
\multiput(872.00,208.17)(15.707,-13.000){2}{\rule{0.312pt}{0.400pt}}
\multiput(889.00,194.92)(0.808,-0.491){17}{\rule{0.740pt}{0.118pt}}
\multiput(889.00,195.17)(14.464,-10.000){2}{\rule{0.370pt}{0.400pt}}
\multiput(905.00,184.93)(1.088,-0.488){13}{\rule{0.950pt}{0.117pt}}
\multiput(905.00,185.17)(15.028,-8.000){2}{\rule{0.475pt}{0.400pt}}
\multiput(922.00,176.93)(1.255,-0.485){11}{\rule{1.071pt}{0.117pt}}
\multiput(922.00,177.17)(14.776,-7.000){2}{\rule{0.536pt}{0.400pt}}
\multiput(939.00,169.94)(2.236,-0.468){5}{\rule{1.700pt}{0.113pt}}
\multiput(939.00,170.17)(12.472,-4.000){2}{\rule{0.850pt}{0.400pt}}
\multiput(955.00,165.95)(3.588,-0.447){3}{\rule{2.367pt}{0.108pt}}
\multiput(955.00,166.17)(12.088,-3.000){2}{\rule{1.183pt}{0.400pt}}
\put(591.0,780.0){\rule[-0.200pt]{4.095pt}{0.400pt}}
\multiput(988.00,164.61)(3.588,0.447){3}{\rule{2.367pt}{0.108pt}}
\multiput(988.00,163.17)(12.088,3.000){2}{\rule{1.183pt}{0.400pt}}
\multiput(1005.00,167.58)(0.732,0.492){19}{\rule{0.682pt}{0.118pt}}
\multiput(1005.00,166.17)(14.585,11.000){2}{\rule{0.341pt}{0.400pt}}
\multiput(1021.00,178.61)(3.588,0.447){3}{\rule{2.367pt}{0.108pt}}
\multiput(1021.00,177.17)(12.088,3.000){2}{\rule{1.183pt}{0.400pt}}
\multiput(1038.58,177.37)(0.494,-0.977){29}{\rule{0.119pt}{0.875pt}}
\multiput(1037.17,179.18)(16.000,-29.184){2}{\rule{0.400pt}{0.438pt}}
\multiput(1054.00,148.92)(0.607,-0.494){25}{\rule{0.586pt}{0.119pt}}
\multiput(1054.00,149.17)(15.784,-14.000){2}{\rule{0.293pt}{0.400pt}}
\multiput(1071.00,134.93)(1.395,-0.482){9}{\rule{1.167pt}{0.116pt}}
\multiput(1071.00,135.17)(13.579,-6.000){2}{\rule{0.583pt}{0.400pt}}
\multiput(1087.00,128.95)(3.588,-0.447){3}{\rule{2.367pt}{0.108pt}}
\multiput(1087.00,129.17)(12.088,-3.000){2}{\rule{1.183pt}{0.400pt}}
\put(1104,125.17){\rule{3.500pt}{0.400pt}}
\multiput(1104.00,126.17)(9.736,-2.000){2}{\rule{1.750pt}{0.400pt}}
\put(1121,123.67){\rule{3.854pt}{0.400pt}}
\multiput(1121.00,124.17)(8.000,-1.000){2}{\rule{1.927pt}{0.400pt}}
\put(972.0,164.0){\rule[-0.200pt]{3.854pt}{0.400pt}}
\put(161,737){\usebox{\plotpoint}}
\put(194,735.67){\rule{4.095pt}{0.400pt}}
\multiput(194.00,736.17)(8.500,-1.000){2}{\rule{2.048pt}{0.400pt}}
\put(161.0,737.0){\rule[-0.200pt]{7.950pt}{0.400pt}}
\put(260,734.67){\rule{4.095pt}{0.400pt}}
\multiput(260.00,735.17)(8.500,-1.000){2}{\rule{2.048pt}{0.400pt}}
\put(211.0,736.0){\rule[-0.200pt]{11.804pt}{0.400pt}}
\put(310,733.67){\rule{3.854pt}{0.400pt}}
\multiput(310.00,734.17)(8.000,-1.000){2}{\rule{1.927pt}{0.400pt}}
\put(326,732.67){\rule{4.095pt}{0.400pt}}
\multiput(326.00,733.17)(8.500,-1.000){2}{\rule{2.048pt}{0.400pt}}
\put(277.0,735.0){\rule[-0.200pt]{7.950pt}{0.400pt}}
\put(360,731.67){\rule{3.854pt}{0.400pt}}
\multiput(360.00,732.17)(8.000,-1.000){2}{\rule{1.927pt}{0.400pt}}
\put(376,730.17){\rule{3.500pt}{0.400pt}}
\multiput(376.00,731.17)(9.736,-2.000){2}{\rule{1.750pt}{0.400pt}}
\put(393,728.67){\rule{3.854pt}{0.400pt}}
\multiput(393.00,729.17)(8.000,-1.000){2}{\rule{1.927pt}{0.400pt}}
\put(409,727.17){\rule{3.500pt}{0.400pt}}
\multiput(409.00,728.17)(9.736,-2.000){2}{\rule{1.750pt}{0.400pt}}
\put(426,725.17){\rule{3.300pt}{0.400pt}}
\multiput(426.00,726.17)(9.151,-2.000){2}{\rule{1.650pt}{0.400pt}}
\multiput(442.00,723.95)(3.588,-0.447){3}{\rule{2.367pt}{0.108pt}}
\multiput(442.00,724.17)(12.088,-3.000){2}{\rule{1.183pt}{0.400pt}}
\multiput(459.00,720.94)(2.236,-0.468){5}{\rule{1.700pt}{0.113pt}}
\multiput(459.00,721.17)(12.472,-4.000){2}{\rule{0.850pt}{0.400pt}}
\multiput(475.00,716.94)(2.382,-0.468){5}{\rule{1.800pt}{0.113pt}}
\multiput(475.00,717.17)(13.264,-4.000){2}{\rule{0.900pt}{0.400pt}}
\multiput(492.00,712.93)(1.712,-0.477){7}{\rule{1.380pt}{0.115pt}}
\multiput(492.00,713.17)(13.136,-5.000){2}{\rule{0.690pt}{0.400pt}}
\multiput(508.00,707.93)(1.255,-0.485){11}{\rule{1.071pt}{0.117pt}}
\multiput(508.00,708.17)(14.776,-7.000){2}{\rule{0.536pt}{0.400pt}}
\multiput(525.00,700.93)(1.022,-0.488){13}{\rule{0.900pt}{0.117pt}}
\multiput(525.00,701.17)(14.132,-8.000){2}{\rule{0.450pt}{0.400pt}}
\multiput(541.00,692.92)(0.860,-0.491){17}{\rule{0.780pt}{0.118pt}}
\multiput(541.00,693.17)(15.381,-10.000){2}{\rule{0.390pt}{0.400pt}}
\multiput(558.00,682.92)(0.655,-0.493){23}{\rule{0.623pt}{0.119pt}}
\multiput(558.00,683.17)(15.707,-13.000){2}{\rule{0.312pt}{0.400pt}}
\multiput(575.00,669.92)(0.531,-0.494){27}{\rule{0.527pt}{0.119pt}}
\multiput(575.00,670.17)(14.907,-15.000){2}{\rule{0.263pt}{0.400pt}}
\multiput(591.58,653.83)(0.495,-0.528){31}{\rule{0.119pt}{0.524pt}}
\multiput(590.17,654.91)(17.000,-16.913){2}{\rule{0.400pt}{0.262pt}}
\multiput(608.58,635.20)(0.494,-0.721){29}{\rule{0.119pt}{0.675pt}}
\multiput(607.17,636.60)(16.000,-21.599){2}{\rule{0.400pt}{0.338pt}}
\multiput(624.58,612.05)(0.495,-0.768){31}{\rule{0.119pt}{0.712pt}}
\multiput(623.17,613.52)(17.000,-24.523){2}{\rule{0.400pt}{0.356pt}}
\multiput(641.58,585.58)(0.494,-0.913){29}{\rule{0.119pt}{0.825pt}}
\multiput(640.17,587.29)(16.000,-27.288){2}{\rule{0.400pt}{0.413pt}}
\multiput(657.58,556.26)(0.495,-1.009){31}{\rule{0.119pt}{0.900pt}}
\multiput(656.17,558.13)(17.000,-32.132){2}{\rule{0.400pt}{0.450pt}}
\multiput(674.58,521.85)(0.494,-1.137){29}{\rule{0.119pt}{1.000pt}}
\multiput(673.17,523.92)(16.000,-33.924){2}{\rule{0.400pt}{0.500pt}}
\multiput(690.58,485.87)(0.495,-1.129){31}{\rule{0.119pt}{0.994pt}}
\multiput(689.17,487.94)(17.000,-35.937){2}{\rule{0.400pt}{0.497pt}}
\multiput(707.58,447.64)(0.494,-1.201){29}{\rule{0.119pt}{1.050pt}}
\multiput(706.17,449.82)(16.000,-35.821){2}{\rule{0.400pt}{0.525pt}}
\multiput(723.58,409.97)(0.495,-1.099){31}{\rule{0.119pt}{0.971pt}}
\multiput(722.17,411.99)(17.000,-34.985){2}{\rule{0.400pt}{0.485pt}}
\multiput(740.58,373.17)(0.495,-1.039){31}{\rule{0.119pt}{0.924pt}}
\multiput(739.17,375.08)(17.000,-33.083){2}{\rule{0.400pt}{0.462pt}}
\multiput(757.58,338.16)(0.494,-1.041){29}{\rule{0.119pt}{0.925pt}}
\multiput(756.17,340.08)(16.000,-31.080){2}{\rule{0.400pt}{0.463pt}}
\multiput(773.58,305.75)(0.495,-0.858){31}{\rule{0.119pt}{0.782pt}}
\multiput(772.17,307.38)(17.000,-27.376){2}{\rule{0.400pt}{0.391pt}}
\multiput(790.58,276.89)(0.494,-0.817){29}{\rule{0.119pt}{0.750pt}}
\multiput(789.17,278.44)(16.000,-24.443){2}{\rule{0.400pt}{0.375pt}}
\multiput(806.58,251.44)(0.495,-0.648){31}{\rule{0.119pt}{0.618pt}}
\multiput(805.17,252.72)(17.000,-20.718){2}{\rule{0.400pt}{0.309pt}}
\multiput(823.58,229.61)(0.494,-0.593){29}{\rule{0.119pt}{0.575pt}}
\multiput(822.17,230.81)(16.000,-17.807){2}{\rule{0.400pt}{0.288pt}}
\multiput(839.00,211.92)(0.497,-0.495){31}{\rule{0.500pt}{0.119pt}}
\multiput(839.00,212.17)(15.962,-17.000){2}{\rule{0.250pt}{0.400pt}}
\multiput(856.00,194.92)(0.616,-0.493){23}{\rule{0.592pt}{0.119pt}}
\multiput(856.00,195.17)(14.771,-13.000){2}{\rule{0.296pt}{0.400pt}}
\multiput(872.00,181.92)(0.712,-0.492){21}{\rule{0.667pt}{0.119pt}}
\multiput(872.00,182.17)(15.616,-12.000){2}{\rule{0.333pt}{0.400pt}}
\multiput(889.00,169.93)(0.902,-0.489){15}{\rule{0.811pt}{0.118pt}}
\multiput(889.00,170.17)(14.316,-9.000){2}{\rule{0.406pt}{0.400pt}}
\multiput(905.00,160.93)(1.088,-0.488){13}{\rule{0.950pt}{0.117pt}}
\multiput(905.00,161.17)(15.028,-8.000){2}{\rule{0.475pt}{0.400pt}}
\multiput(922.00,152.93)(1.255,-0.485){11}{\rule{1.071pt}{0.117pt}}
\multiput(922.00,153.17)(14.776,-7.000){2}{\rule{0.536pt}{0.400pt}}
\multiput(939.00,145.93)(1.712,-0.477){7}{\rule{1.380pt}{0.115pt}}
\multiput(939.00,146.17)(13.136,-5.000){2}{\rule{0.690pt}{0.400pt}}
\multiput(955.00,140.94)(2.382,-0.468){5}{\rule{1.800pt}{0.113pt}}
\multiput(955.00,141.17)(13.264,-4.000){2}{\rule{0.900pt}{0.400pt}}
\multiput(972.00,136.95)(3.365,-0.447){3}{\rule{2.233pt}{0.108pt}}
\multiput(972.00,137.17)(11.365,-3.000){2}{\rule{1.117pt}{0.400pt}}
\multiput(988.00,133.95)(3.588,-0.447){3}{\rule{2.367pt}{0.108pt}}
\multiput(988.00,134.17)(12.088,-3.000){2}{\rule{1.183pt}{0.400pt}}
\put(1005,130.17){\rule{3.300pt}{0.400pt}}
\multiput(1005.00,131.17)(9.151,-2.000){2}{\rule{1.650pt}{0.400pt}}
\put(1021,128.17){\rule{3.500pt}{0.400pt}}
\multiput(1021.00,129.17)(9.736,-2.000){2}{\rule{1.750pt}{0.400pt}}
\put(1038,126.67){\rule{3.854pt}{0.400pt}}
\multiput(1038.00,127.17)(8.000,-1.000){2}{\rule{1.927pt}{0.400pt}}
\put(1054,125.67){\rule{4.095pt}{0.400pt}}
\multiput(1054.00,126.17)(8.500,-1.000){2}{\rule{2.048pt}{0.400pt}}
\put(1071,124.67){\rule{3.854pt}{0.400pt}}
\multiput(1071.00,125.17)(8.000,-1.000){2}{\rule{1.927pt}{0.400pt}}
\put(1087,123.67){\rule{4.095pt}{0.400pt}}
\multiput(1087.00,124.17)(8.500,-1.000){2}{\rule{2.048pt}{0.400pt}}
\put(343.0,733.0){\rule[-0.200pt]{4.095pt}{0.400pt}}
\put(1104.0,124.0){\rule[-0.200pt]{7.950pt}{0.400pt}}
\end{picture}
\vspace{0.5cm}
\caption{Illustrative plot for matter effects on ${\cal J}$
associated with neutrinos ($+A$) and antineutrinos ($-A$),
where $\Delta m^2_{21} = 5\cdot 10^{-5} ~ {\rm eV}^2$ and 
$\Delta m^2_{31} = 3\cdot 10^{-3} ~ {\rm eV}^2$ have been input.}
\end{figure}

We observe that matter effects can be significant for the elements in 
the first and the second columns of $V$, if $A \geq 10^{-5} ~ {\rm eV}^2$. 
In comparison, the magnitudes of $|V_{e3}|$, $|V_{\mu 3}|$ and
$|V_{\tau 3}|$ may be drastically enhanced or suppressed only for
$A > 10^{-3} ~ {\rm eV}^2$. The neutrinos are relatively more sensitive to
the matter effects than the antineutrinos.

The magnitude of ${\cal J}_{\rm m}$ decreases, when the matter
effect becomes significant (e.g., $A \geq 10^{-4} ~{\rm eV}^2$).
However, this does not imply that the CP- or T-violating asymmetries 
in realistic long-baseline neutrino oscillations would be smaller than
their values in vacuum. Large matter effects can significantly modify  
the frequencies of neutrino oscillations and thus enhance or
suppress the genuine signals of CP or T violation.

If the earth-induced matter effects can well be controlled, it is 
possible to recast the fundamental flavor 
mixing matrix $V$ from a variety of measurements of 
neutrino oscillations. Such a goal is expected to be
reached in the neutrino factories \cite{Factory,Barger00}.

I would like to thank H. Fritzsch and A. Thomas for supporting my p
articipation in this interesting conference.

\end{document}